\documentclass[aps,pre,twocolumn,showpacs,floatfix,superscriptaddress,tightenlines,amsmath,amssymb,yhmath]{revtex4-2}
\usepackage[utf8]{inputenc}
\usepackage[T1]{fontenc}
\usepackage{hyperref}
\usepackage[dvipsnames,x11names]{xcolor}
\pagecolor{white}
\usepackage[final]{graphicx}
\usepackage{bm}
\usepackage{float}
\usepackage{enumerate, enumitem}
\usepackage[normalem]{ulem}
\usepackage{xcolor}
\usepackage{multirow}
\usepackage{hyperref}
\hypersetup{
    colorlinks = true,
    urlcolor   = blue,
    citecolor  = green,
}

\definecolor{RoyalBlue}{HTML}{007399}
\definecolor{Seagreen}{HTML}{009999}
\definecolor{crimsonred}{HTML}{990000}
\definecolor{Purple}{HTML}{66032C}
\definecolor{Yellow}{HTML}{DFC98A}
\definecolor{darkgreen}{HTML}{006400}
\definecolor{darkyellow}{HTML}{C9A227}

\providecommand\bnabla{\boldsymbol{\nabla}}

\newcommand{\REM}[1]{{{}}}

\DeclareMathAlphabet\mathbfcal{OMS}{cmsy}{b}{n}

\def \dd{{\rm d}}

\def\bu {\boldsymbol{u}}

\def \bx {\boldsymbol{x}}

\def\i{{\rm i}}

\def \Ga  {\mbox{Ga}}
\def \At  {\mbox{At}}
\def \Bo  {\mbox{Bo}}

\begin{document}

\title{Gappy Reconstruction of Bubbly Flows by Guided Diffusion Models}
\author{Hridey Narula}
\email{hrideynarula@tifrh.res.in}
\affiliation{Tata Institute of Fundamental Research, Gopanpally, Hyderabad 500046, India}
\author{Tianyi Li}
\author{Michele Buzzicotti}
\author{Luca Biferale}
\affiliation{Department of Physics and INFN, University of Rome “Tor Vergata”, Via della Ricerca Scientifica 1, Rome, 00133, Italy}
\author{Prasad Perlekar}
\email{perlekar@tifrh.res.in}
\affiliation{Tata Institute of Fundamental Research, Gopanpally, Hyderabad 500046, India}

\date{\today}

\begin{abstract}
Experiments in multiphase flows are often limited in their ability to simultaneously obtain velocity measurements in different phases.
At the same time, flow reconstruction from phase-limited measurements is a challenging problem due to the substantially different velocity statistics across the phases.
We address this problem for buoyancy-driven bubbly flows in the pseudo-turbulence regime by using a guided diffusion model.
We train the model using two-dimensional slices of the velocity field extracted from fully resolved three-dimensional direct numerical simulations.
The model generates physically realistic velocity fields both unconditionally and when conditioned on the surrounding liquid flow.
The reconstructed bubble-phase velocity field accurately reproduces key statistical features of the flow.
We further show that a simple stacking procedure for adjacent two-dimensional slices enables a reasonable reconstruction of the three-dimensional flow inside a bubble.
These results establish the potential of diffusion models to serve as generative priors for three-dimensional turbulent multiphase flows, opening a route toward the reconstruction of unobserved or experimentally inaccessible velocity fields from sparse, partial, or phase-limited measurements.
\end{abstract}
\maketitle

\section{Introduction}
Flows generated by rising columns of bubbles are encountered in a wide range of natural and industrial settings \citep{clift2005bubbles,Balchandar_2010}.
At moderate gas volume fractions ($0.5$–$10\%$), bubbles are homogeneously distributed, and wakes associated with individual bubbles interact and superimpose, giving rise to complex multiscale flow structures commonly referred to as pseudo-turbulence (PT) or bubble-induced agitation \citep{LanceBataille1991,Risso_2018,ChaoSun_2020}. 
Although the flow generated by a single bubble strongly depends on the density and viscosity contrast between the bubble and liquid phases \citep{weber1981,clift2005bubbles, Tripathi2014,Tripathi_2015}, the collective dynamics of a large number of bubbles appears to have some universal properties \citep{LanceBataille1991,ZENIT_2013,Prakash_2016,Scarbolo_2016,Elghobashi_2019,Pandey_Ramadugu_Perlekar_2020,Innocenti_Jaccod_Popinet_Chibbaro_2021,Mangani_2022,PandeyMitraPerlekar2023,RAMIREZ2024104860,Ravisankar_Zenit_2024,ma2025,Fabien_2025,Narula_2026jfm}.
These studies suggest that the complex interplay of inertia, surface tension, and viscous dissipation leads to PT, which is characterized by the Galilei number ($\Ga$)--the ratio of buoyancy to viscous forces.
Similar to canonical turbulence, a statistical approach is used to understand the nature of velocity fluctuations and energy transfers in bubbly flows.
At moderate Galilei numbers ($\Ga\lesssim1000$), the energy $E(k)$ contained in a wavenumber shell $k$ has been shown to obey the scaling $k^{-3}$ \citep{LanceBataille1991,Risso_2011,Prakash_2016,Sun_2017,Amoura_Besnaci_Risso_Roig_2017,Risso_2018,Pandey_Ramadugu_Perlekar_2020}. More recently, studies have shown the emergence of the Kolmogorov scaling $k^{-5/3}$ at sufficiently large Galilei numbers ($\Ga\gtrsim1000$) \citep{PandeyMitraPerlekar2023,ma2025}.

In spite of recent advances, the multiphase and multiscale nature of the bubbly flow makes their numerical and experimental investigations extremely challenging. 
Accurate fully resolved direct numerical simulations are often limited to tens or hundreds of bubbles, and are computationally prohibitive~\citep{ESMAEELI_TRYGGVASON_1999,Pandey_Ramadugu_Perlekar_2020,Innocenti_Jaccod_Popinet_Chibbaro_2021,PandeyMitraPerlekar2023}.
Similarly, experiments can typically only monitor liquid velocity fluctuations using either probes, such as a hot-wire~\citep{Mercado_2007,ALMERAS2019316}, or Lagrangian particle tracking measurements~\citep{HESSENKEMPER_2018,Ma_Hessenkemper_Lucas_Bragg_2022}.
Although machine-learning-assisted flow visualization techniques have been employed for precise tracking of individual bubbles in a swarm~\citep{POLETAEV_2020,HAAS_2020,CERQUEIRA_2021,Kim2021,CUI_2022,HESSENKEMPER_2022,WANG_2023,Hessenkemper_2024,kucuk_2025}, resolving the flow dynamics within the bubbles remains difficult.

Attempts to use AI/ML models for generating realistic flow fields in bubbly flows have largely been confined to the case of either an isolated bubble or flows at small Galilei numbers.
\citet{Zhai_2022} have previously employed training data obtained from two-dimensional ($2d$) simulations of pressure-driven channel flows with up to $60$ bubbles at small Reynolds numbers $(\lesssim 0.01)$ to  show that physics-informed neural networks (PINNs) can successfully predict flow fields.

However, for flows with even moderately large Galilei numbers, flow reconstruction remains challenging because not only does the flow resemble canonical turbulence, but also the velocity statistics in the liquid and bubble phases are substantially different~\citep{RIBOUX_2010,Mercado_2010,ROGHAIR_2011,Pandey_Ramadugu_Perlekar_2020,Innocenti_Jaccod_Popinet_Chibbaro_2021}.\\
Recently, generative diffusion models have emerged as effective tools for accurately representing and reconstructing turbulent flow fields in a variety of contexts, including: (a) producing synthetic tracer and inertial-particle trajectories in turbulent and active-matter flows \citep{Li_nmi2024,Li_ijmpf2024,Jha_2026}; and (b) completing missing segments of (gappy) Lagrangian trajectories \citep{Li_comm2025} and reconstructing Eulerian flow fields \citep{SHU_2023,Li_atmos2024,Trepat_2026}.

In the present work, we have developed a generative diffusion model to unconditionally generate two-dimensional slices of three-dimensional buoyancy-driven bubbly flows, and subsequently reconstruct the flow inside the bubbles conditioned on the ambient liquid flow information for a moderately large Galilei number ($\Ga=600$).\\
The remainder of the paper is organized as follows: in Section~\ref{sec:dns_details}, we first introduce the governing Navier–Stokes equations for two-phase flows and describe the numerical methodologies employed to generate the training dataset for bubbly flows.
Subsequently, in Section~\ref{sec:diffusion_model}, we present a concise overview of the theoretical framework underlying the diffusion model.
Finally, in Section~\ref{sec:flow_inside_bubble}, we report the main results concerning the two-dimensional reconstruction of the flow within bubbles at a moderate Galilei number $\Ga = 600$, along with a discussion of generalization to different Galilei numbers as well as a naive three-dimensional reconstruction procedure.
\section{Dataset Details\label{sec:dns_details}}
\subsection{Numerical Methods}
We model the dynamics of bubbly flows in a tri-periodic cubic domain of volume $V$ using the one-fluid formulation \citep{ESMAEELI_TRYGGVASON_1998,BUNNER_TRYGGVASON_2002}. The incompressible Navier-Stokes equations for the hydrodynamic velocity field $\bu$ in the two phases are:
\begin{subequations}
\begin{align}
  \rho(\phi) D_t \bu &= -\bnabla P + \mu \nabla^2 \bu + \boldsymbol{F}^g + \boldsymbol{F}^\sigma, \\
 D_t \phi &= 0,~{\rm and}~\bnabla   \cdot \bu= 0.
\end{align}
\label{eq:nse}
\end{subequations}
Here, $D_t=\partial_t + \bu \cdot \bnabla$ is the material derivative, $\phi$ is an indicator function that takes the value $\phi=0$ ($1$) in the liquid (bubble) phase, the density field $\rho(\phi)=\rho_B \phi + \rho_L(1-\phi)$ where $\rho_L$ and $\rho_B$ are the densities of the liquid and bubble phases, respectively. $P$ is the pressure, $\mu$ is the dynamic viscosity, assumed to be the same in the two phases, $\boldsymbol{F}^g=(\rho-\rho_a) g \hat{z}$ is the buoyancy force, where the average density $\rho_a\equiv 1/V \int \rho\, {\rm d}{\boldsymbol{x}}$.
The surface tension force  $\boldsymbol{F}^\sigma=\sigma\kappa\boldsymbol{n}$, where $\sigma$ is the surface tension coefficient, $\kappa$ is the local interface curvature, and $\boldsymbol{n}$ is the normal vector to the interface.

In previous studies, we have shown that several statistical properties of the flow are insensitive to the value of the Atwood number, $\At \equiv (\rho_L - \rho_B)/(\rho_L + \rho_B)$ \citep{Pandey_Ramadugu_Perlekar_2020,Pandey_Mitra_Perlekar_2022,PandeyMitraPerlekar2023}. 
Consequently, for simplicity, we  use $\At = 0.04$ and invoke the Boussinesq approximation wherein density variations are neglected on the left-hand side of Eq.~\eqref{eq:nse} ($\rho(\phi) \approx \rho_L$), while the buoyancy forcing term reduces to $\boldsymbol{F}^g \approx (\rho_B - \rho_L)\,\phi\,g\,\hat{\boldsymbol{z}}$ \citep{Pandey_Mitra_Perlekar_2022}.   
\subsection{Direct Numerical Simulations (DNS)}
\begin{figure*}
\raisebox{5.5mm}{\includegraphics[width=0.3\linewidth]{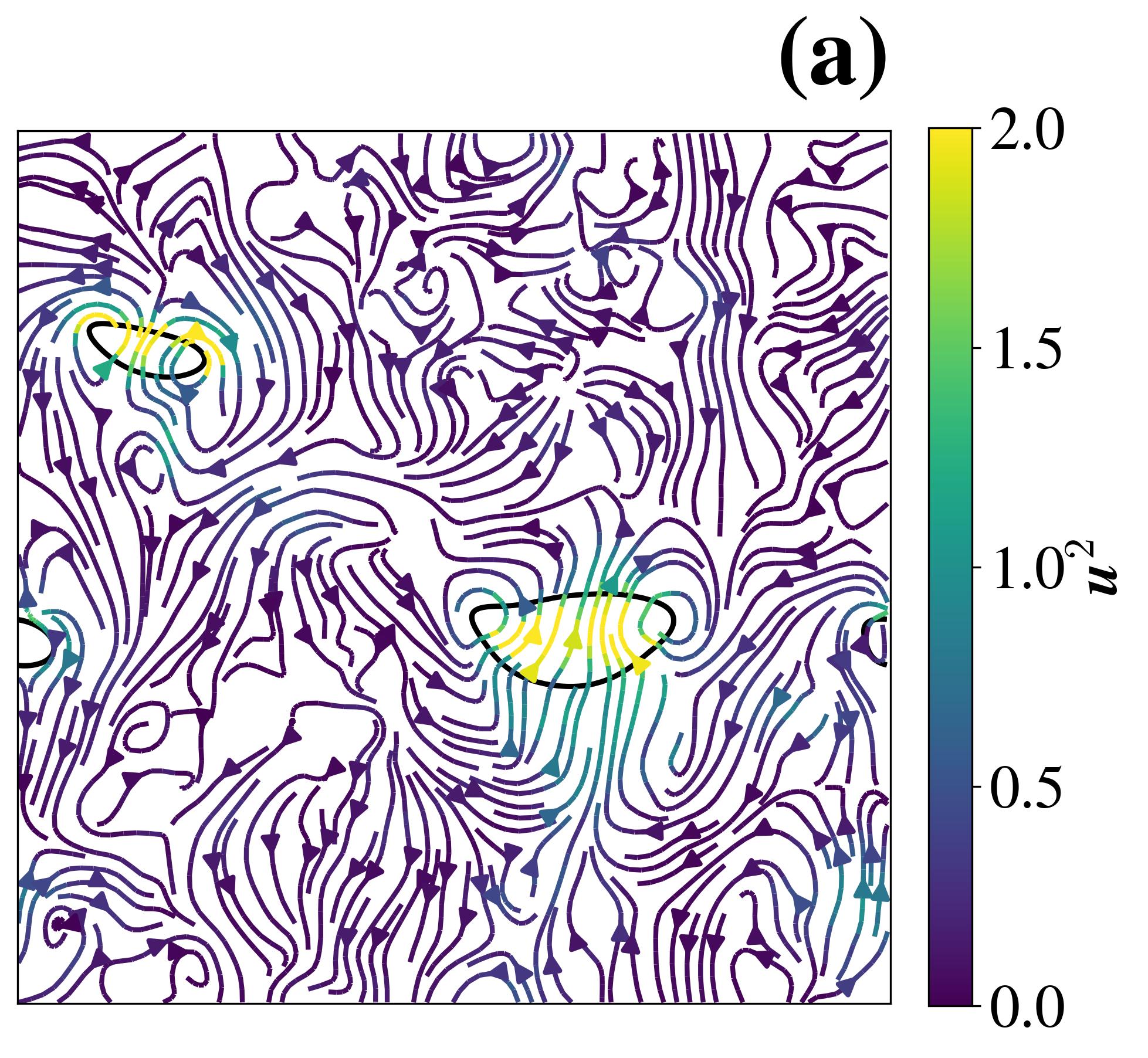}}
  \includegraphics[width=0.65\linewidth]{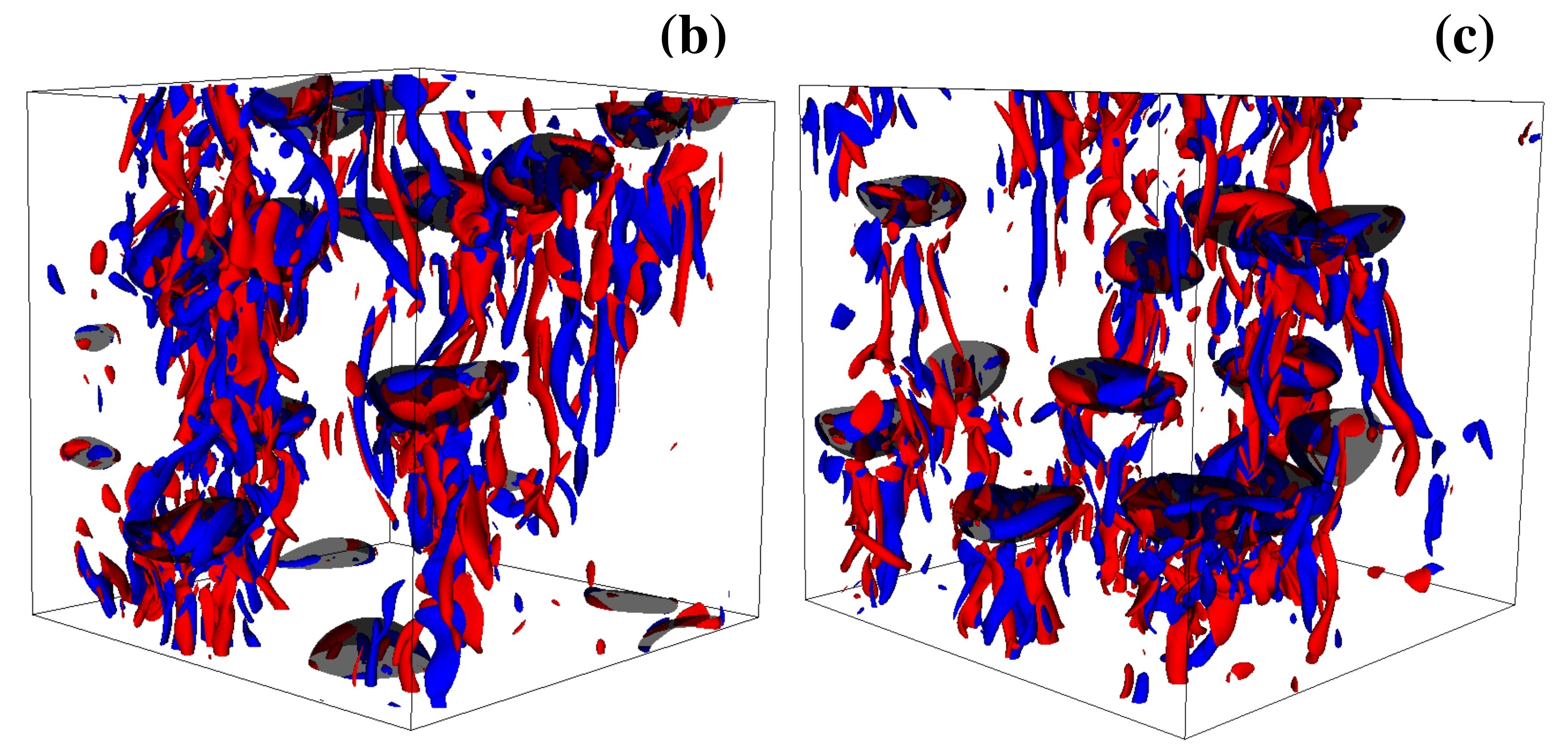}
\caption{\label{fig:bubble_vis}
(a) Streamlines of the instantaneous velocity field in the plane $z=L/2$. (b, c) Iso-surfaces of the $z$-component of vorticity, $\omega_z=(\nabla \times \bu)\cdot\hat{z}$, computed at the levels $\omega_z=\pm 3\langle \omega_z^2\rangle^{1/2}$ ($+$ in red and $-$ in blue), and superposed on the bubble interface (the $\phi=1/2$ isocontour, grey) for two statistically independent instants.}
\end{figure*}

We perform DNS of Eq. \eqref{eq:nse} on a triperiodic cubic domain with side length $L=2\pi$ discretized with $N^3=512^3$ equispaced collocation points.
The velocity field is evolved using a pseudo-spectral method \citep{Can88}, while the indicator function is evolved with a front-tracking algorithm \citep{univerdi_1992,TRYGGVASON_2001,Aniszewski2021}.
Time integration employs a second-order Adams–Bashforth scheme. The details of the numerical implementation and the comparison with bubbly flow experiments \citep{Prakash_2016,Mercado_2010,ROGHAIR_2011} are discussed in \citet{Pandey_Ramadugu_Perlekar_2020,Pandey_Mitra_Perlekar_2022}.
The simulation domain is initialized with $12$ randomly placed spherical bubbles of equal diameter $D$ with the gas volume fraction $1/V\int\phi\,\dd\bx=3.2\%$. 
The other parameters are chosen so that the dimensionless numbers describing the system, the Galilei number $\Ga\equiv \sqrt{\rho_L (\rho_L-\rho_B)g D^3}/\mu\approx 600$ and the Bond number $\Bo\equiv(\rho_L-\rho_B)g D^2/\sigma \approx 1.75$ are chosen to be comparable to the typical experimental values \citep{LanceBataille1991, Prakash_2016,Risso_2018,ma2025}.
We perform DNS with these parameters to generate three independent flow  realizations, each initialized with different initial bubble positions.
Table~\ref{tab:dataset} summarizes the parameters used in our simulations.
In Fig.~\ref{fig:bubble_vis} we show instantaneous realizations of the bubbly flow  to highlight the complex spatio-temporal flow structures generated by bubble wakes and their interactions.

\begin{table*}
  \begin{center}
\def~{\hphantom{0}}
\begin{tabular}{lcccccc}
\textbf{Run} & $T/\tau_\eta$ & $\delta t/\tau_\eta$ &
\textbf{Number of Snapshots} & \textbf{Number of Samples} & \textbf{Used for}\\[3pt]
\hline
\tt R1  & 273 & 2.25 & 122 & 10,630 & Training \\
\tt R2  & 160 & 2.25 &  72 &  6,314 & Training \\
\tt R3  & 167 & 2.25 &  75 &  6,407 & -- \\
\quad\raisebox{.2ex}{\(\vdash\)}\tt R3a
       & -- & -- & -- & 300   & Validation \\
\quad\raisebox{.2ex}{\(\vdash\)}\tt R3b
       & -- & -- & -- & 3,200 & Testing \\
\end{tabular}
  \caption{Details about the DNS datasets used. All the runs have identical $\Ga\approx600$ and $\Bo\approx1.75$, but correspond to different initial bubble positions.
  The number of bubbles is 12 (volume fraction $\approx 3.2\%$).
  The Kolmogorov timescale is
  $\tau_\eta=\sqrt{\nu/\varepsilon}$, where $\nu$ is the liquid kinematic viscosity and $\varepsilon$ is the energy dissipation rate.
   All simulations are with $512^3$ collocation points, with $\eta/\Delta x\approx1.59$ where $\eta$ is the Kolmogorov lengthscale, $\eta=(\nu^3/\varepsilon)^{1/4}$ and $\Delta x$ is the grid spacing in real space.
  $T$ is the total duration of the statistically steady state and $\delta t$ is the interval between successive snapshots.
  Number of snapshots are the total three-dimensional snapshots and number of samples are the total two-dimensional samples obtained from the same.
  {\tt R3a} and {\tt R3b} are mutually exclusive datasets obtained from run {\tt R3}. {\tt R3a} is used to find the optimal guidance strength and {\tt R3b} for final testing. See section \ref{subsec_data_acquire} for details of the data acquisition.
  }
  \label{tab:dataset}
  \end{center}
\end{table*}

\begin{figure*}
  \includegraphics[width=0.98\linewidth]{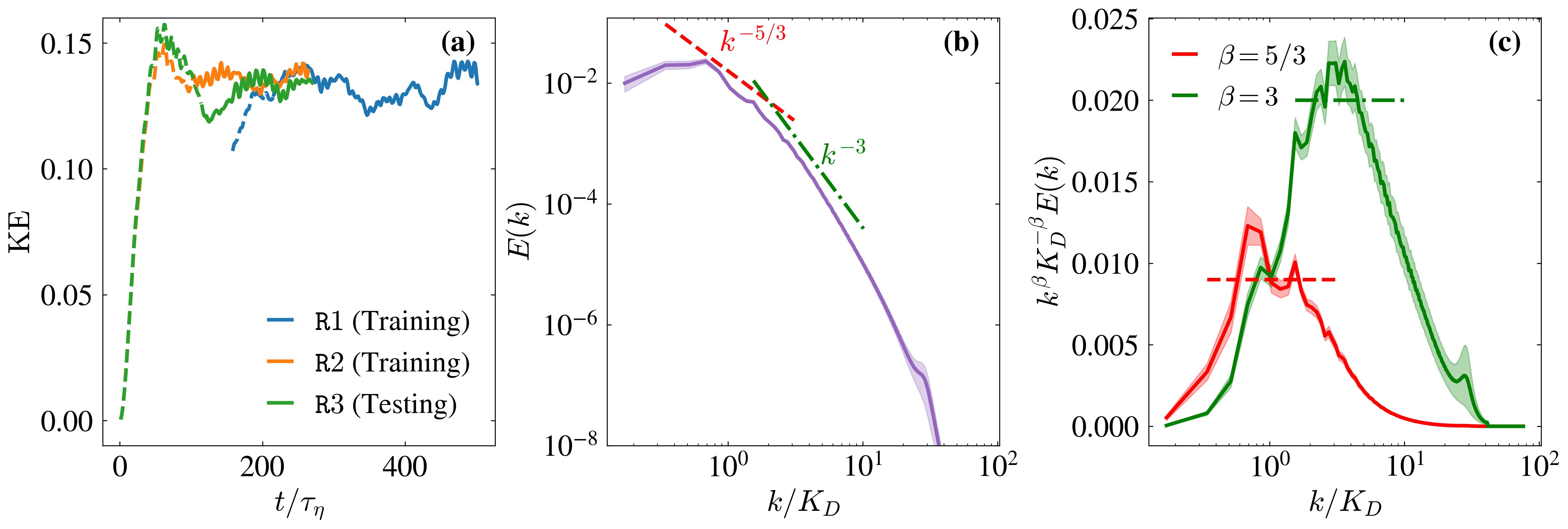}
\caption{\label{fig:dataset_specs}(a) Time evolution of the kinetic energy for three different initial bubble configurations. The initial transient phase is shown with dashed lines, while solid lines denote the statistically stationary regime ($t>100\tau_\eta$). (b) Steady-state energy spectrum $E(k)$ versus $k/K_D$ averaged over the three runs; the shaded band denotes the $1\sigma$ spread, where $K_D=2\pi/D$ and $\sigma$ is the standard deviation. The dashed and dash-dotted reference lines indicate the Kolmogorov ($k^{-5/3}$) and pseudo-turbulence ($k^{-3}$) scalings, respectively. (c) Energy spectrum shown in compensated form using the Kolmogorov and pseudo-turbulence scalings; shading again indicates $1\sigma$ variations.}

\end{figure*}
The evolution of the average kinetic energy ${\rm KE} = (1/2V) \int \rho \bu^2 \dd \bx$ is shown in Fig.~\ref{fig:dataset_specs}a.
A statistically stationary regime is achieved for $t>100\tau_\eta$, where $\tau_\eta\equiv \sqrt{\nu/\varepsilon}$ is the Kolmogorov timescale with $\nu$ being the liquid-phase kinematic viscosity and $\varepsilon$ being the energy dissipation rate.
Fig.~\ref{fig:dataset_specs}b displays the corresponding steady-state energy spectrum defined as:
\begin{equation}
E(k) = \frac{1}{2} \sum\limits_{k-1/2\le\kappa\le k+1/2}|\widehat{\bu}(\boldsymbol{\kappa})|^2,\nonumber    
\end{equation}
where $\widehat{\bu}$ denotes the Fourier transform of the velocity field.
Energy injection occurs primarily near the wavenumber $k\approx K_D$, where $K_D=2\pi/D$.
We also indicate the Kolmogorov scaling range $k^{-5/3}$ for $K_D<k<k_\eta$ \citep{PandeyMitraPerlekar2023,ma2025}, as well as the pseudo-turbulent scaling range $k^{-3}$ for $k>k_\eta$ \citep{LanceBataille1991,Prakash_2016,Risso_2018,Pandey_Ramadugu_Perlekar_2020,Innocenti_Jaccod_Popinet_Chibbaro_2021} and the compensated spectra in Fig.~\ref{fig:dataset_specs}c.

\section{Diffusion Models for Flow Generation and Reconstruction\label{sec:diffusion_model}}
\subsection{Unconditional Generative Model}
We briefly summarize the diffusion-model framework used in this work.
Diffusion models are a class of generative probabilistic models that learn a high-dimensional data distribution through a prescribed \textit{forward noising process} and a learned \textit{reverse (backward) denoising process} \citep{Sohl_2015,Ho_2020}.
Let $\mathbf{X}_0$ denote a representative sample from the training dataset (details about the dataset are provided in Section~\ref{sec:2d_flow_inside_bubble}).
The objective of the diffusion model is to learn a generative distribution $p_\theta(\mathbf{X}_0)$ that approximates the underlying training-data (target) distribution $q(\mathbf{X}_0)$, where $\theta$ denotes the neural-network parameters.
Once trained, new samples can be generated starting from white Gaussian noise and iteratively applying the learned reverse process.

In the forward process, the clean flow configuration $\mathbf{X}_0$ is gradually transformed into a sequence of noisy states
$\mathbf{X}_1,\ldots,\mathbf{X}_T$ through a Gaussian Markov chain (note that $t$ here is the discrete noising step),
\begin{equation}\label{eq:markov_chain_forward}
q(\mathbf{X}_t|\mathbf{X}_{t-1}) = \mathcal{N}\left(
\sqrt{1-\beta_t}\,\mathbf{X}_{t-1},
\beta_t \mathbf{I}
\right),
\end{equation}
where $\beta_t$ is the prescribed noise variance at diffusion step $t$.
Defining $\alpha_t=1-\beta_t$ and
$\bar{\alpha}_t=\prod_{s=1}^{t}\alpha_s$, one obtains the closed-form expression
\begin{equation}
q(\mathbf{X}_t|\mathbf{X}_0)
=
\mathcal{N}\left(
\sqrt{\bar{\alpha}_t}\,\mathbf{X}_0,
(1-\bar{\alpha}_t)\mathbf{I}
\right),
\end{equation}
or equivalently
\begin{equation}\label{eq:forward_sample_closed}
\mathbf{X}_t
=
\sqrt{\bar{\alpha}_t}\,\mathbf{X}_0
+
\sqrt{1-\bar{\alpha}_t}\,\boldsymbol{\epsilon},
\qquad
\boldsymbol{\epsilon}\sim\mathcal{N}(\mathbf{0},\mathbf{I}).
\end{equation}
For sufficiently large $T$, $\mathbf{X}_T$ is approximately distributed according to a standard Gaussian distribution.

The backward process approximates the inverse of the forward noising process by denoising from $\mathbf{X}_t$ to $\mathbf{X}_{t-1}$ at each diffusion step.
It models the reverse transition probability as
\begin{equation}\label{eq:reverse_transition}
p_\theta(\mathbf{X}_{t-1}|\mathbf{X}_t)
=
\mathcal{N}\left(
\boldsymbol{\mu}_\theta(\mathbf{X}_t,t),
\boldsymbol{\Sigma}_t
\right),
\end{equation}
where $\boldsymbol{\mu}_\theta$ is the learned reverse mean.
Following the standard DDPM (Denoising Diffusion Probabilistic Model) formulation \citep{Ho_2020}, we fix the reverse variance according to the prescribed noise schedule, $\boldsymbol{\Sigma}_t=\beta_t\mathbf{I}$.
Instead of directly predicting $\boldsymbol{\mu}_\theta$, a neural network is trained to predict the noise component $\boldsymbol{\epsilon}_\theta(\mathbf{X}_t,t)$ in Eq. \eqref{eq:forward_sample_closed}.
The corresponding estimate of the clean sample (denoted using a hat) is
\begin{equation}\label{eq:x0_estimate}
\hat{\mathbf{X}}_0(\mathbf{X}_t,t)
=
\frac{
\mathbf{X}_t-\sqrt{1-\bar{\alpha}_t}\,
\boldsymbol{\epsilon}_\theta(\mathbf{X}_t,t)
}{
\sqrt{\bar{\alpha}_t}
}.
\end{equation}
The reverse mean is then written in terms of the predicted noise as \citep{Ho_2020}
\begin{equation}\label{eq:reverse_mean}
\boldsymbol{\mu}_\theta(\mathbf{X}_t,t)
=
\frac{1}{\sqrt{\alpha_t}}
\left[
\mathbf{X}_t
-
\frac{\beta_t}{\sqrt{1-\bar{\alpha}_t}}
\boldsymbol{\epsilon}_\theta(\mathbf{X}_t,t)
\right].
\end{equation}
Starting from $\mathbf{X}_T\sim\mathcal{N}(\mathbf{0},\mathbf{I})$, repeated application of Eq.~\eqref{eq:reverse_transition} generates a sample from the learned distribution $p_\theta(\mathbf{X}_0)$.

The model is trained using the standard DDPM noise-prediction objective, which corresponds to a simplified variational bound on the expected data log-likelihood $\mathbb{E}_{q(\mathbf{X}_0)}[\log p_\theta(\mathbf{X}_0)]$ \citep{Ho_2020}.
At each training iteration, a batch of clean samples $\mathbf{X}_0$ is drawn from the training dataset; for each data sample, a diffusion step $t$ and a noise realization $\boldsymbol{\epsilon}$ are sampled, and the corresponding noisy sample $\mathbf{X}_t$ is constructed using Eq.~\eqref{eq:forward_sample_closed}.
The model parameters are optimized by minimizing
\begin{equation}\label{eq:ddpm_loss}
    L =
    \mathbb{E}_{\mathbf{X}_0,t,\boldsymbol{\epsilon}}
    \left[\left\|
    \boldsymbol{\epsilon}-\boldsymbol{\epsilon}_\theta(\mathbf{X}_t,t)
    \right\|^2\right],
\end{equation}
where $\boldsymbol{\epsilon}\sim\mathcal{N}(\mathbf{0},\mathbf{I})$ is the noise used to generate $\mathbf{X}_t$.
The loss curves shown below report the mini-batch value of \eqref{eq:ddpm_loss} during training.
Additional implementation details, including the neural-network architecture (U-net~\cite{Ronneberger_2015}), diffusion noise schedule, and training setup, are provided in the supplementary material~\cite{SM_self2026} (which includes Refs.~\cite{Vaswani_2023,Loshchilov_2019,ZENIT_KOCH_SANGANI_2001,RENSEN_LUTHER_LOHSE_2005}); in particular, we use a physically motivated noise schedule to better preserve small-scale fidelity in the generated flow fields.

\subsection{Training-Free Conditional Generation}\label{subsec:tfg_dps}
The unconditional diffusion model has learned the prior distribution $p_\theta(\mathbf{X}_0)$ over the clean training dataset.
We can generate new (unconditioned) samples from this distribution using Eq.~\eqref{eq:reverse_transition}.
However, for guided generation we want to generate a new sample conditioned on the given information in the measurement region $\mathcal{M}$.
Therefore, we want the posterior distribution $p_\theta(\mathbf{X}_0|\mathbf{X}_\mathcal{M})$ where $\mathbf{X}_\mathcal{M}$ is the measured portion of the full image.

In the present work, we use a training-free guidance scheme~\citep{ye_2024} so that we can use the same learned prior distribution to generate samples conditioned on the measurement region.
This is achieved by using diffusion posterior sampling (DPS) \citep{Chung2024}, which has been previously shown to be effective in a variety of chaotic systems \citep{Du_2024,LiZeyu_2024,Li_comm2025}. 
In this approach, the reverse process (Eq. \eqref{eq:reverse_transition}) is modified to include a guidance force in the measured region,
\begin{equation}\label{eq:backward_sample_conditional}
    \mathbf{X}_{t-1} = \mu_\theta(\mathbf{X}_t, t) + \sqrt{\beta_t}\mathbf{Z} -\zeta_t \nabla_{\mathbf{X}_t} \| \mathbf{X}_\mathcal{M}-\mathbf{\hat{X}}_0\|^2_\mathcal{M},
\end{equation}
where $\mathbf{Z}\sim\mathcal{N}(\mathbf{0},\mathbf{I})$, $\mathbf{\hat{X}}_0$ is the clean image predicted by the model (using the noisy image $\mathbf{X}_t$ at step $t$, see Eq.\eqref{eq:x0_estimate}) and the norm is calculated only in the measurement region $\mathcal{M}$.
The coefficient $\zeta_t\equiv \zeta/ \| \mathbf{X}_\mathcal{M}-\mathbf{\hat{X}}_0\|$ where the guidance strength (GS) $\zeta$ is a hyperparameter of the guidance protocol \citep{Chung2024}.
The code for both training the model and performing the reconstruction can be found in Ref. \cite{tianyi_CDMlagr}.
In the next section, we use this guided diffusion model framework for reconstruction of the bubble-phase velocity field.
\section{Flow Reconstruction Inside the Bubble phase \label{sec:flow_inside_bubble}}
We now use the diffusion-model architecture described in the previous section for reconstruction of velocity fields inside the bubble. 
Although the gas volume fraction is relatively low (approximately $3.2\%$), flow reconstruction inside bubbles remains highly challenging, since the velocity field is inhomogeneous and its statistical properties differ substantially between the bubble and liquid phases~\citep{Mercado_2007,ROGHAIR_2011,Pandey_Ramadugu_Perlekar_2020}.

Experimental investigations typically provide measurements of two-dimensional ($2d$) flow slices \citep{Ma_Hessenkemper_Lucas_Bragg_2022} or one-dimensional ($1d$) time series \citep{Prakash_2016,LanceBataille1991}.
Consequently, in the present study, we apply diffusion models to reconstruct the velocity field inside the bubble phase, using two-dimensional ($2d$) flow configurations extracted from the DNS data (Table~\ref{tab:dataset}).
In section~\ref{sec:3d_reconstruction}, we demonstrate that, for a prescribed bubble shape, the three-dimensional flow structures inside the bubble can be recovered by assembling the reconstructed $2d$ flow fields. 
We have also used diffusion models for the simpler problem of reconstruction of Eulerian (one-dimensional) time signals; details are provided in the supplementary material to this paper~\cite{SM_self2026}.
\subsection{Two-dimensional Reconstruction of Flow Inside Bubble\label{sec:2d_flow_inside_bubble}}
\subsubsection{Data acquisition\label{subsec_data_acquire}}
The two-dimensional ($2d$) training and testing datasets are constructed from the $3d$ steady-state realizations of the bubbly flows (runs {\tt R1-R3}).
Each dataset is composed of $2d$ slices ($xz$ and $yz$ planes aligned with the gravity direction) of the flow field, and every slice contains at least one bubble. 
To avoid strong correlations, we enforce a minimum separation of $10$ grid points between any two consecutive slices taken from the same $3d$ snapshot.
The $3d$ snapshots are separated by at least $2.25\tau_\eta$.
Furthermore, we apply random spatial shifts to individual slices and shuffle them to destroy any residual spatio-temporal correlations.
For computational feasibility, the chosen planes are subsequently coarse-grained from $512^2$ to $256^2$ grid points using a sharp spectral filter.

We arrange the data in four channels, ${\bf X} = (u_h, u_z, u_p, \phi)$, where $u_h$ and $u_z$ are the horizontal and vertical (in-plane) velocity components, respectively, and $u_p$ is the out-of-plane velocity component. 
To ensure that all four channels have comparable variance, we rescale the velocity field so that ${\bu}\in[-1,1]$ and the indicator field $\phi\in[-1/4,1/4]$ \citep{Li_comm2025}.

\subsubsection{Model training and reconstruction}
\begin{figure}
  \includegraphics[width=0.85\linewidth]{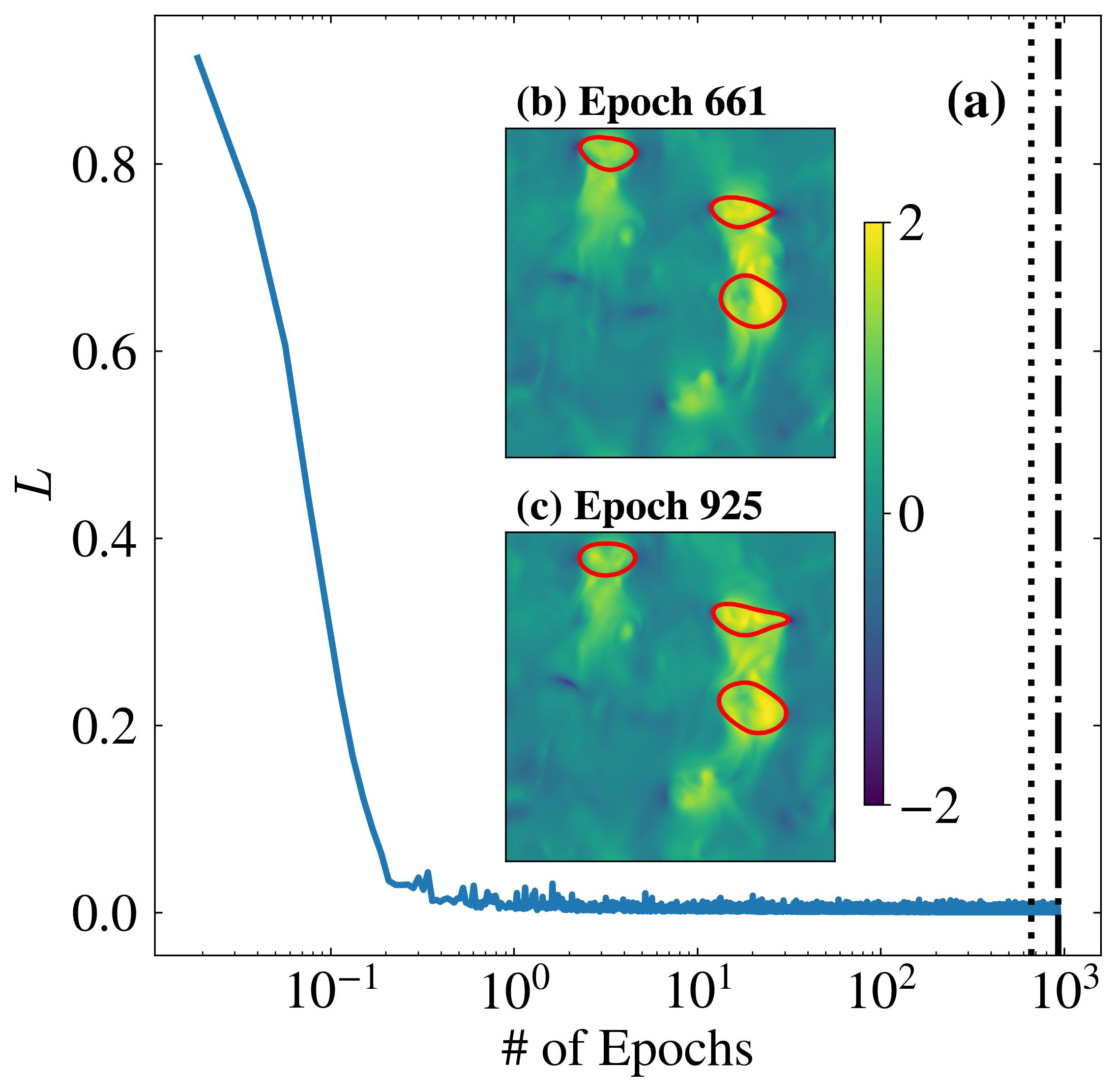}
\caption{\label{fig:training_loss}(a) Loss function $L$ plotted as a function of the number of epochs.
The dot and dash-dot lines indicate epoch number $661$ and $925$ respectively.
Inset: (b, c) Pseudo-color plots of the vertical component of the velocity $u_z$ obtained from the model after training for $661$ (b) and $925$ (c) epochs respectively. The red lines denote the location of the bubble interface.}
\end{figure}

\begin{figure}
  \includegraphics[width=0.9\linewidth]{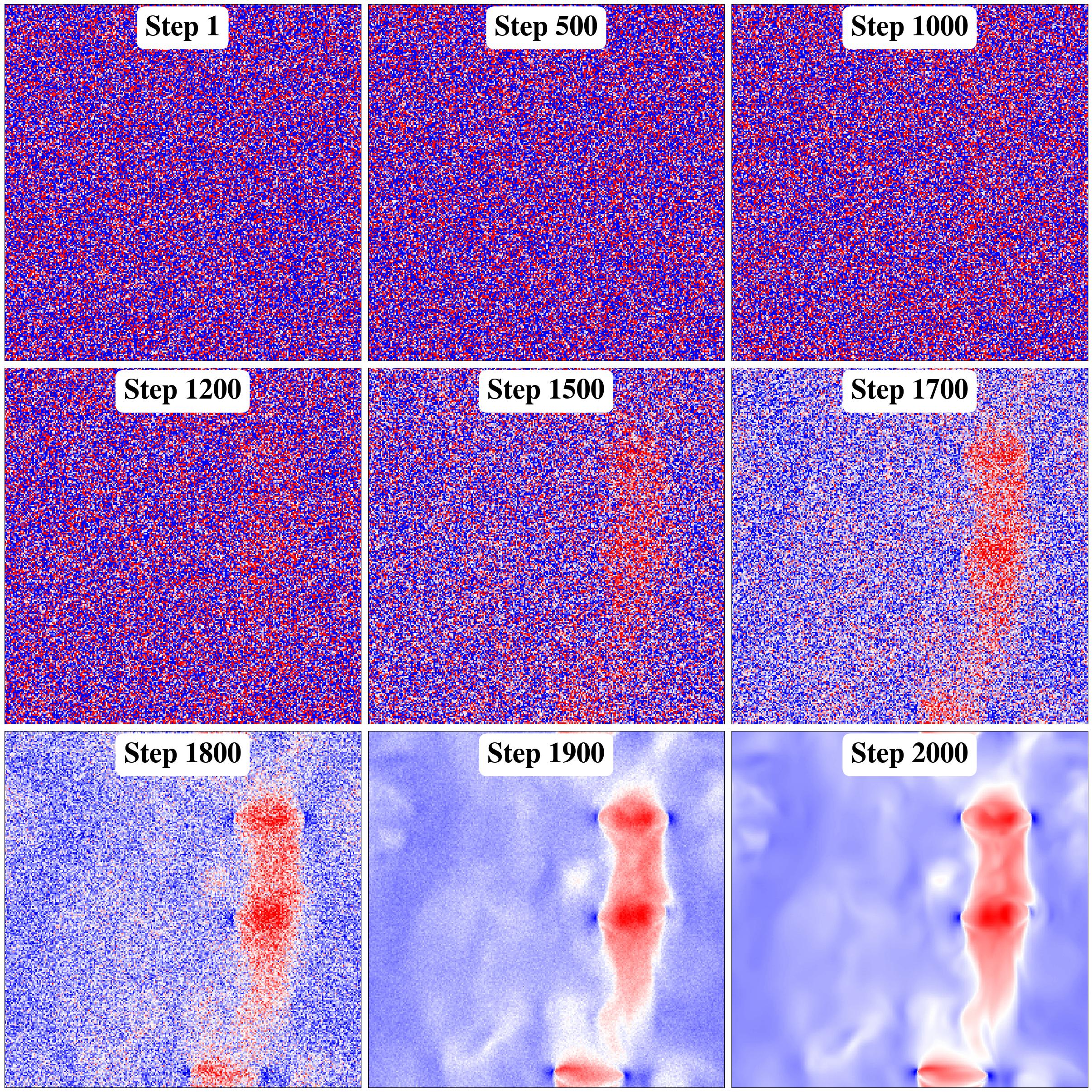}\\
  \includegraphics[width=0.9\linewidth]{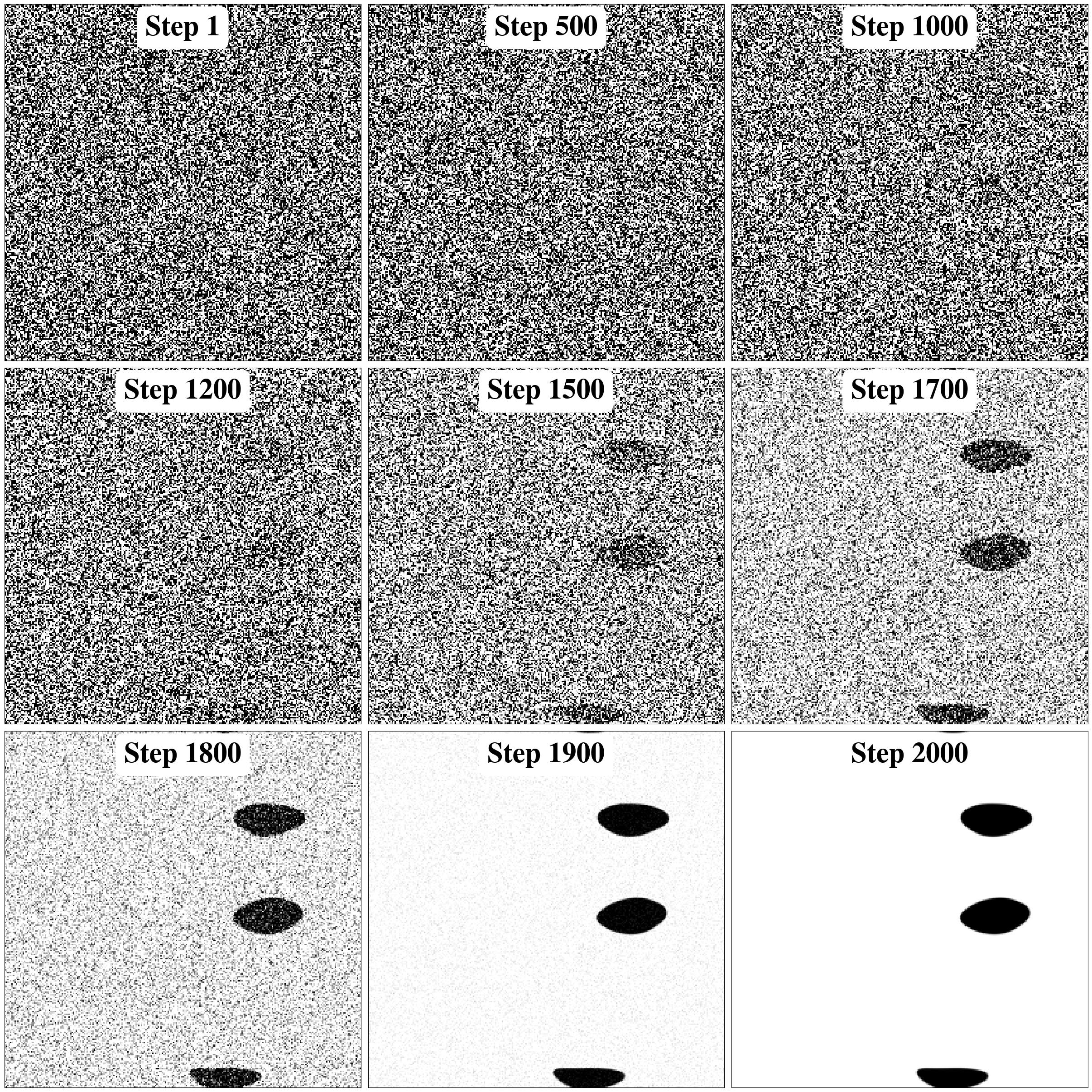}
\caption{\label{fig:guidancefree_denoise} Pseudo-color plot of the vertical velocity field $u_z$ (top) and the gray-scale plot of the indicator function $\phi$ (bottom) evaluated at various stages of the denoising process during sample generation. Top panel: Red (blue) indicates regions of large positive (negative) velocity. Bottom panel: Black (white) denotes the bubble phase $\phi>1/2$ (liquid phase $\phi<1/2$). 
The model starts by considering pure Gaussian noise at step 1 and incrementally denoises it following the backward sampling process (Eq.~\eqref{eq:reverse_transition}).}
\end{figure}

The training set comprises a total of 16,944 slices (runs ${\tt R1}$--${\tt R2}$). We train the model on eight NVIDIA A100 GPUs for about $96$ hours with a batch size of four per GPU. 
Each epoch is defined as $16944/32 \approx 529$ iterations and the model is trained for $925$ epochs.
The plot of the loss function $L$ with respect to the number of epochs in Fig.~\ref{fig:training_loss}a indicates that the training process reaches convergence at approximately $100$ epochs. 
To further substantiate this observation, we present pseudo-color visualizations of the vertical component of the velocity field $u_z$ obtained from a model trained for $661$ epochs (Fig.~\ref{fig:training_loss}b) and from the same model trained for $925$ epochs (Fig.~\ref{fig:training_loss}c). 
The resulting velocity fields are nearly indistinguishable, supporting the conclusion that the training procedure has converged.
In all subsequent analyses, we use the model trained for $925$ epochs ($4.9\times 10^{5}$ iterations).
A representative visualization of the velocity field $u_z$ and the indicator function $\phi$ during various stages of generation (denoising) using our diffusion model is shown in Fig.~\ref{fig:guidancefree_denoise}.

\begin{figure}
\centering 
  \includegraphics[width=0.95\linewidth]{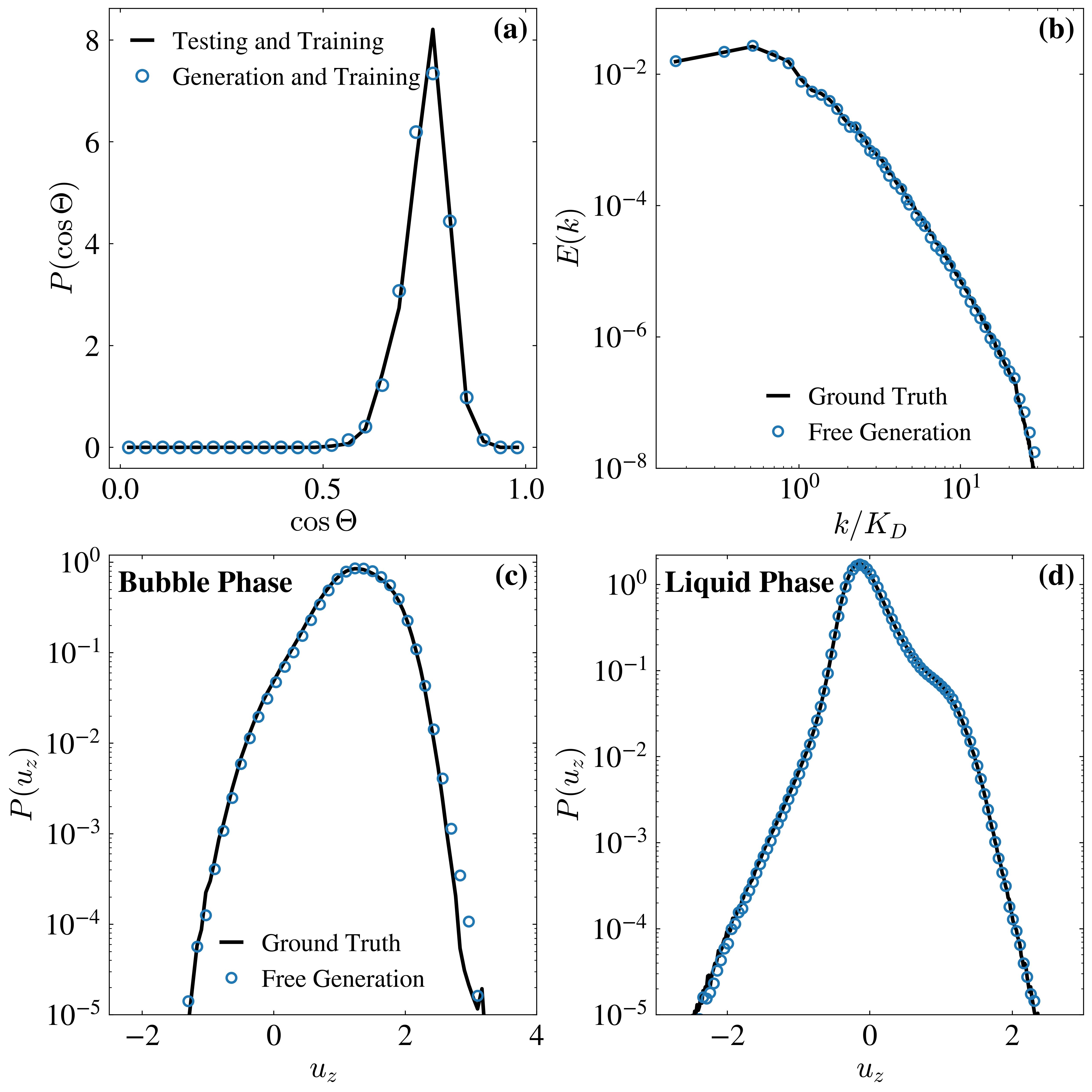}\\
\caption{\label{fig:guidancefree} (a) Plot showing excellent agreement between the PDF of the cosine similarity $P(\cos(\Theta))$ obtained from (i) the testing data and the closest training data, and (ii) the generated data and the closest training data.
(b) The energy spectrum $E(k)$ versus $k$, (c,d) PDF of the vertical velocity $u_z$ in the bubble phase (c) and the liquid phase (d). In (b-d) we use 3,200 slices for both the ground truth (dataset {\tt R3b}) and the free (unconstrained) generation.}
\end{figure}

Diffusion models, similar to other deep learning architectures, are prone to memorizing their training samples~\citep{gu_2025}.
To assess the degree of memorization, we consider the cosine similarity between two fields, say $A$ and $B$, defined as  $\cos(\Theta) = \langle A B \rangle/\sqrt{\langle A^2 \rangle \langle B^2 \rangle}$,
where the angular brackets denote spatial averaging.
We generate $10^3$ unconditioned flow realizations from the model, and for each generated sample determine the closest counterpart in the training set.
This is done by minimizing the mean square difference between the generated realization and the training data for the bubble indicator function $\phi$.
In Fig.~\ref{fig:guidancefree}a, we plot the probability density function (PDF) of cosine similarity $P(\cos(\Theta))$ between (i) $u_z$ obtained from the testing data and its training counterpart, and (ii) $u_z$  generated from the model and its training counterpart.
These distributions are nearly identical, indicating that the velocity field generated by the diffusion model is statistically as similar to the DNS data as is an independent DNS run of the flow with the same parameters.
We thus conclude that any potential memorization effects are not significant in the present analysis.

Next, we present a comparison of various velocity statistics computed from 3,200 slices of both the test dataset (ground truth, run {\tt R3b}) and the generated dataset.
In Fig.~\ref{fig:guidancefree}b, we show the energy spectrum $E(k)$ (obtained by shell-averaging the two-dimensional spectrum of the in-plane velocity field $(u_h,u_z)$) versus $k$, while Fig.~\ref{fig:guidancefree}c,d show the PDF of $u_z$ in the bubble and liquid phases.
In all cases, the agreement between the ground truth and the generated data is excellent, indicating that the model faithfully captures the underlying velocity field statistics in both phases.

\subsubsection{Flow Reconstruction Inside Bubbles}
\begin{figure}
  \includegraphics[width=0.99\linewidth]{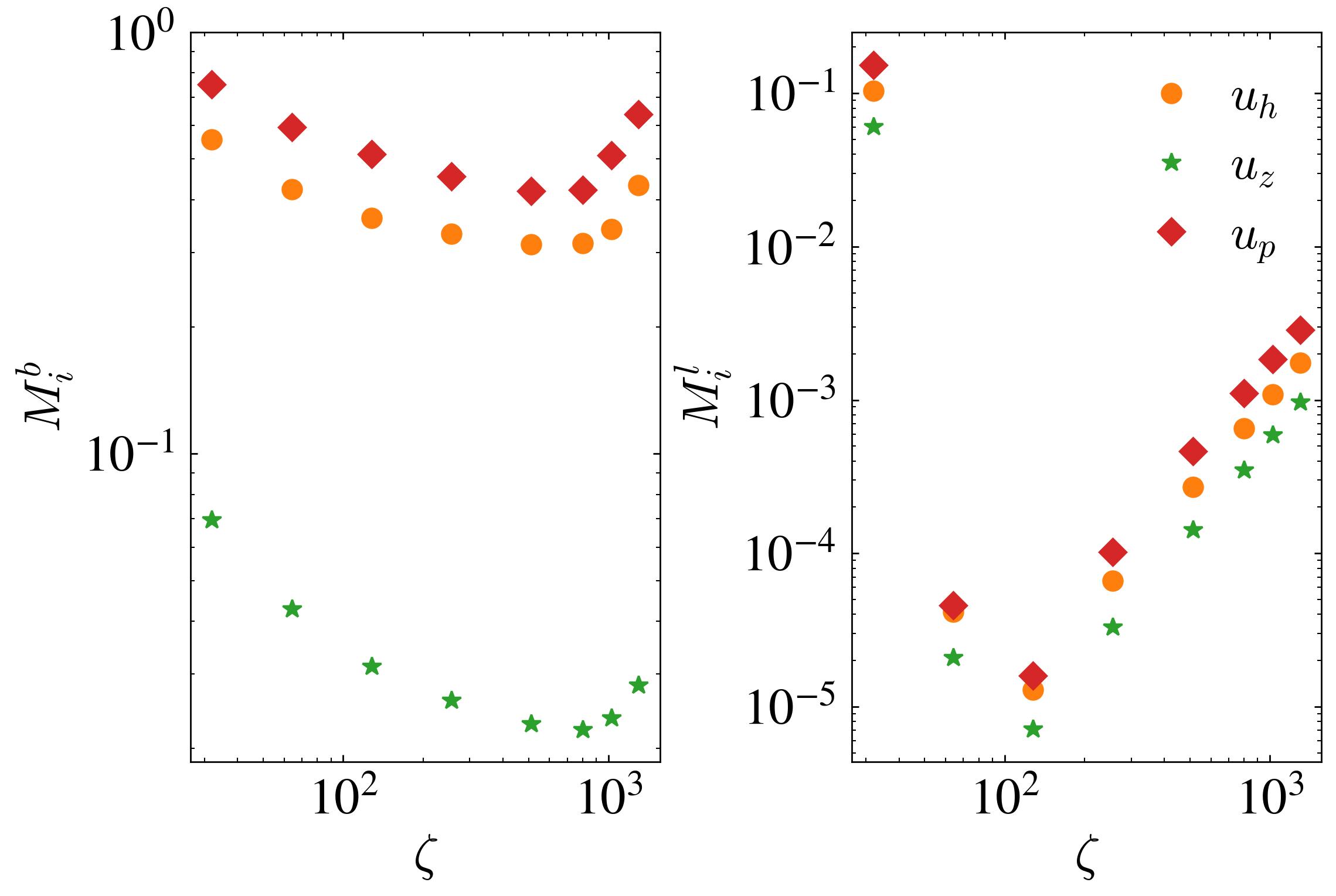}
\caption{\label{fig:optimal_2d_guidance} Normalized mean square error (NMSE) for the different components of velocity: horizontal ($u_h$), vertical ($u_z$), and out-of-plane ($u_p$)—computed between the ground truth and reconstructed samples with varying guidance strength for the bubble phase $M_i^b$ (left) and the liquid phase $M_i^l$ (right).
The mean square error is computed over $300$ slices from run {\tt R3a} that are not included in the final test set.
We select $800$ as the optimal guidance strength, since it corresponds to the minimum value of $M_i^b$ for the three velocity components.}
\end{figure}
\begin{figure*}
\centering
  \includegraphics[scale=0.38]{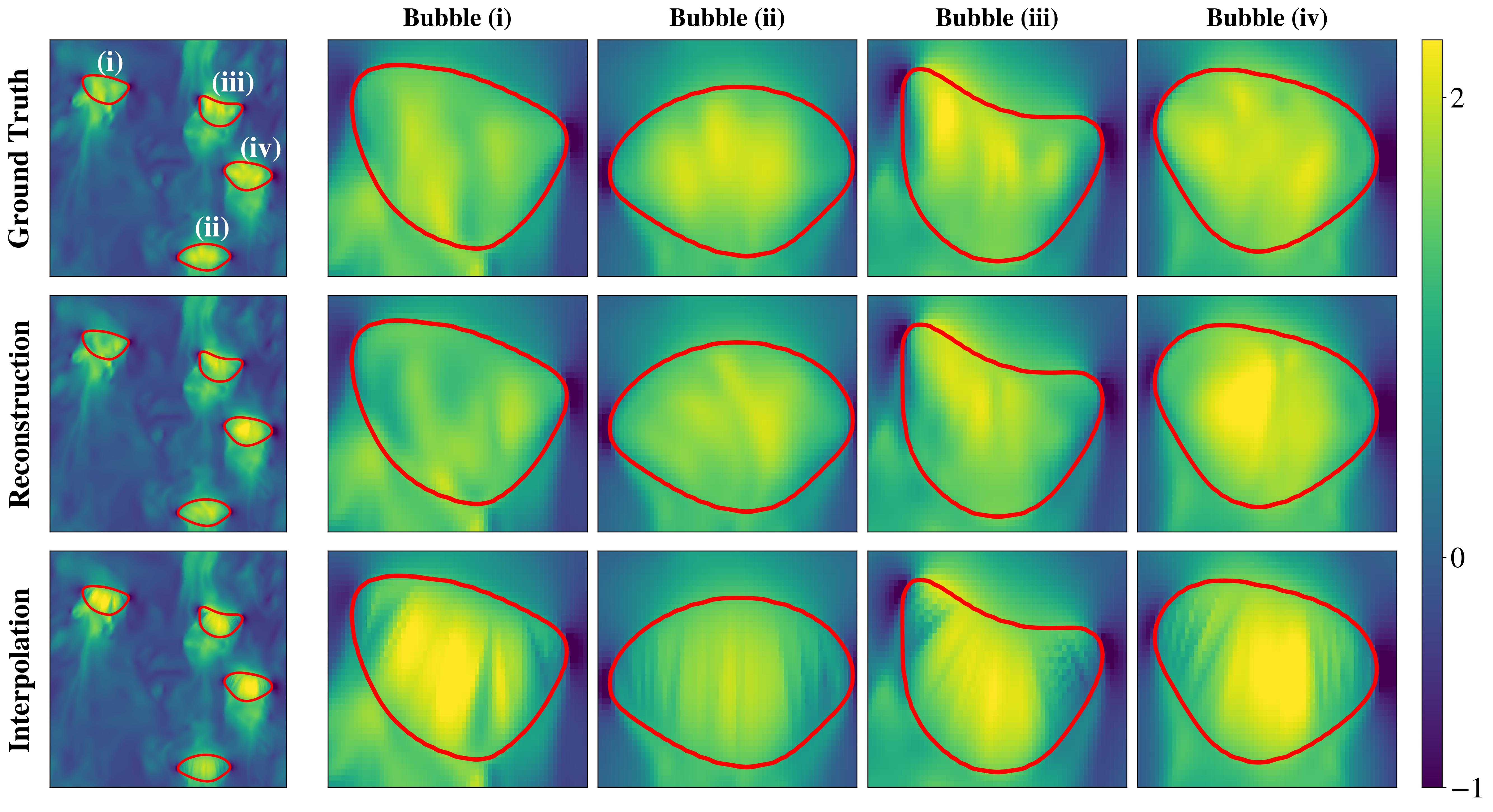}
\caption{A sample $2d$ slice of the vertical velocity field $u_z$ (leftmost column): ground truth (top row), reconstruction using the diffusion model (middle row) and filling using interpolation (bottom row).
On the right are the zoomed-in plots of the velocity field around each bubble.
The red lines demarcate bubble-liquid interface.}\label{fig:reconstruction_visual2d}
\end{figure*}
\begin{figure*}
\centering
  \includegraphics[width=0.92\linewidth]{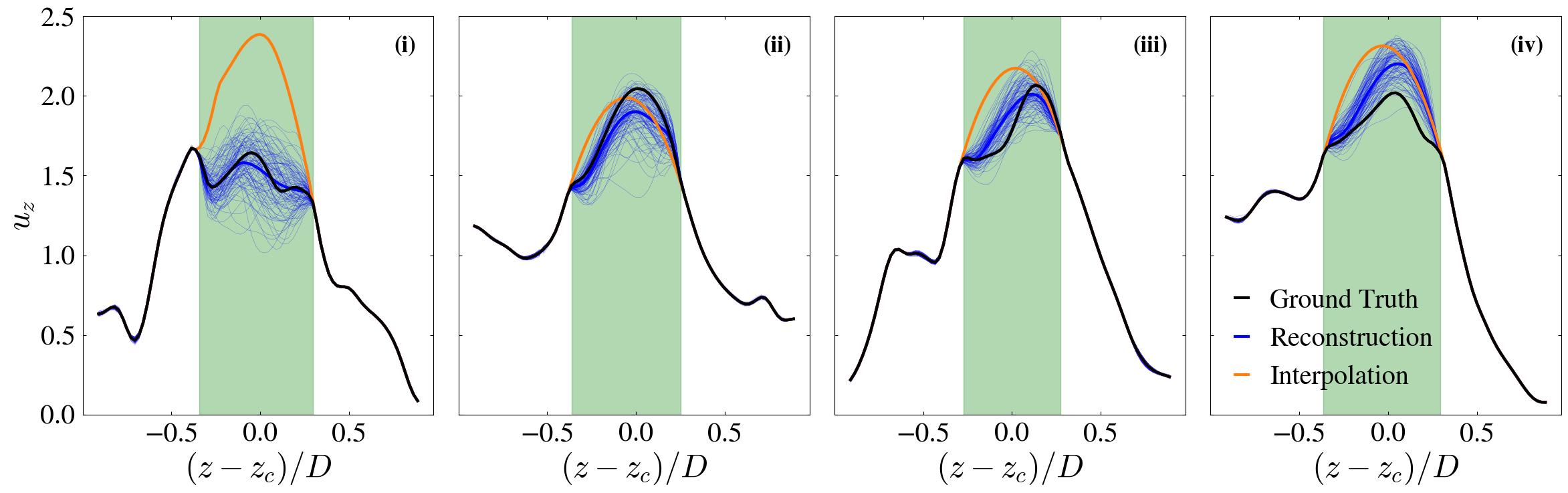}\\
  \includegraphics[width=0.92\linewidth]{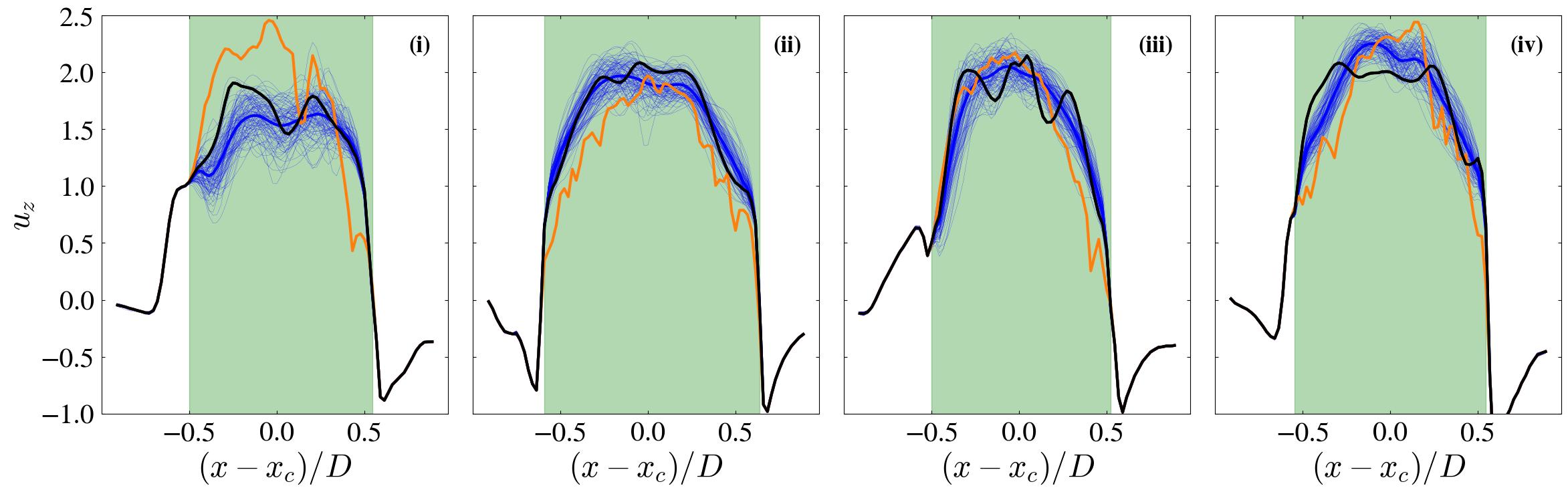}
\caption{One-dimensional cuts of $u_z$ through the center of bubble along the vertical (top) and horizontal (bottom) directions for various reconstructions of the four bubbles in Fig.~\ref{fig:reconstruction_visual2d}.
\textit{Interpolation} refers to $2d$ cubic interpolation using SciPy's griddata functionality.
The shaded region indicates the location of the bubbles. The bubble centers are $(x_c,z_c)$, and $D$ is the bubble diameter.}\label{fig:various_realizations_1dcut}
\end{figure*}
\begin{figure*}
\centering
  \includegraphics[width=0.83\linewidth]{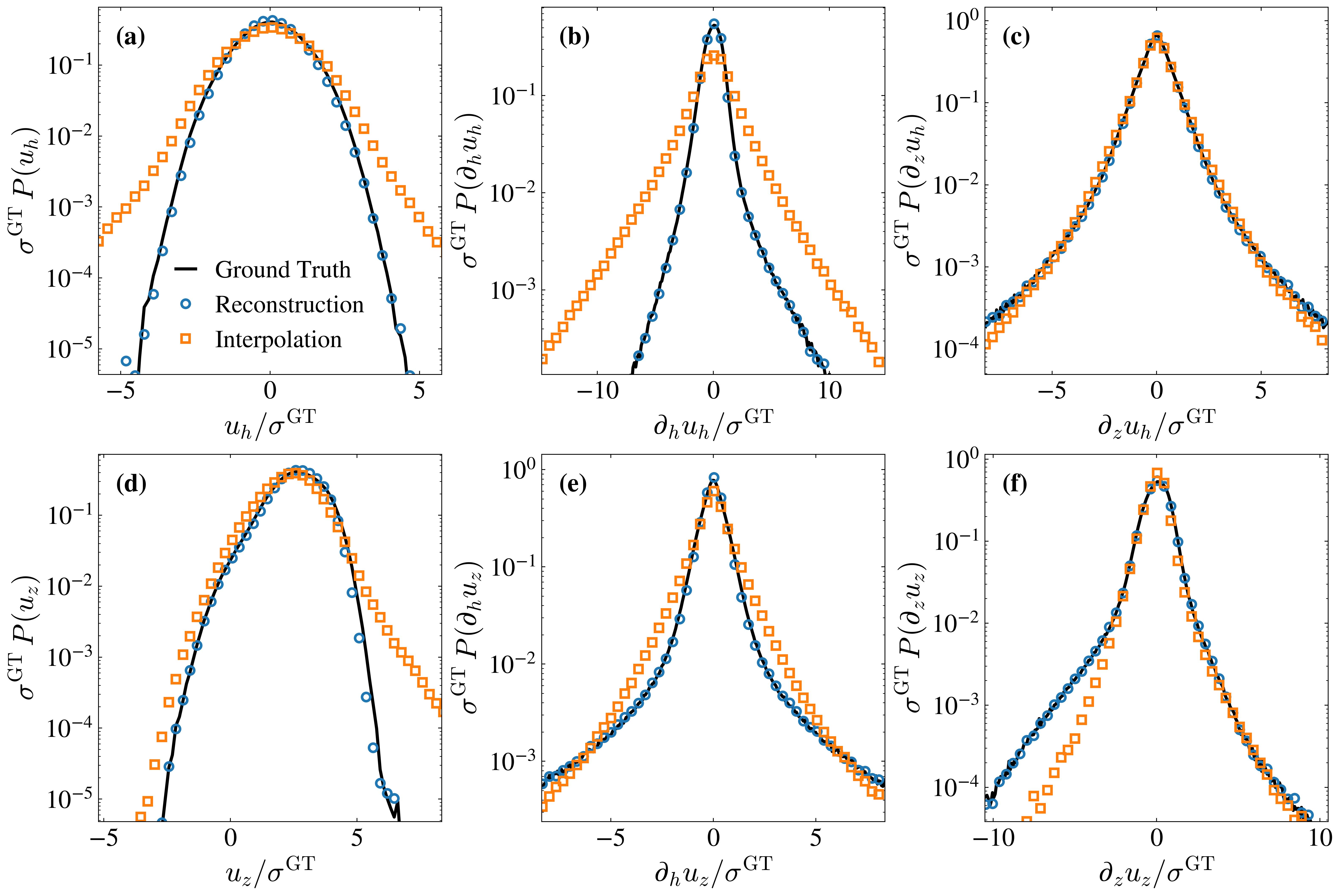}
\caption{PDF of horizontal velocity (a) and its derivatives (b, c) and PDF of vertical velocity (d) and its derivatives (e, f) for the ground truth, reconstruction and interpolation inside the bubble phase.
All the PDFs are standardized by the standard deviation of the ground-truth data, denoted by $\sigma^{\rm GT}$.
\label{fig:pdf_all_bubble_master}}
\end{figure*}
\begin{figure*}
\centering
  \includegraphics[width=0.83\linewidth]{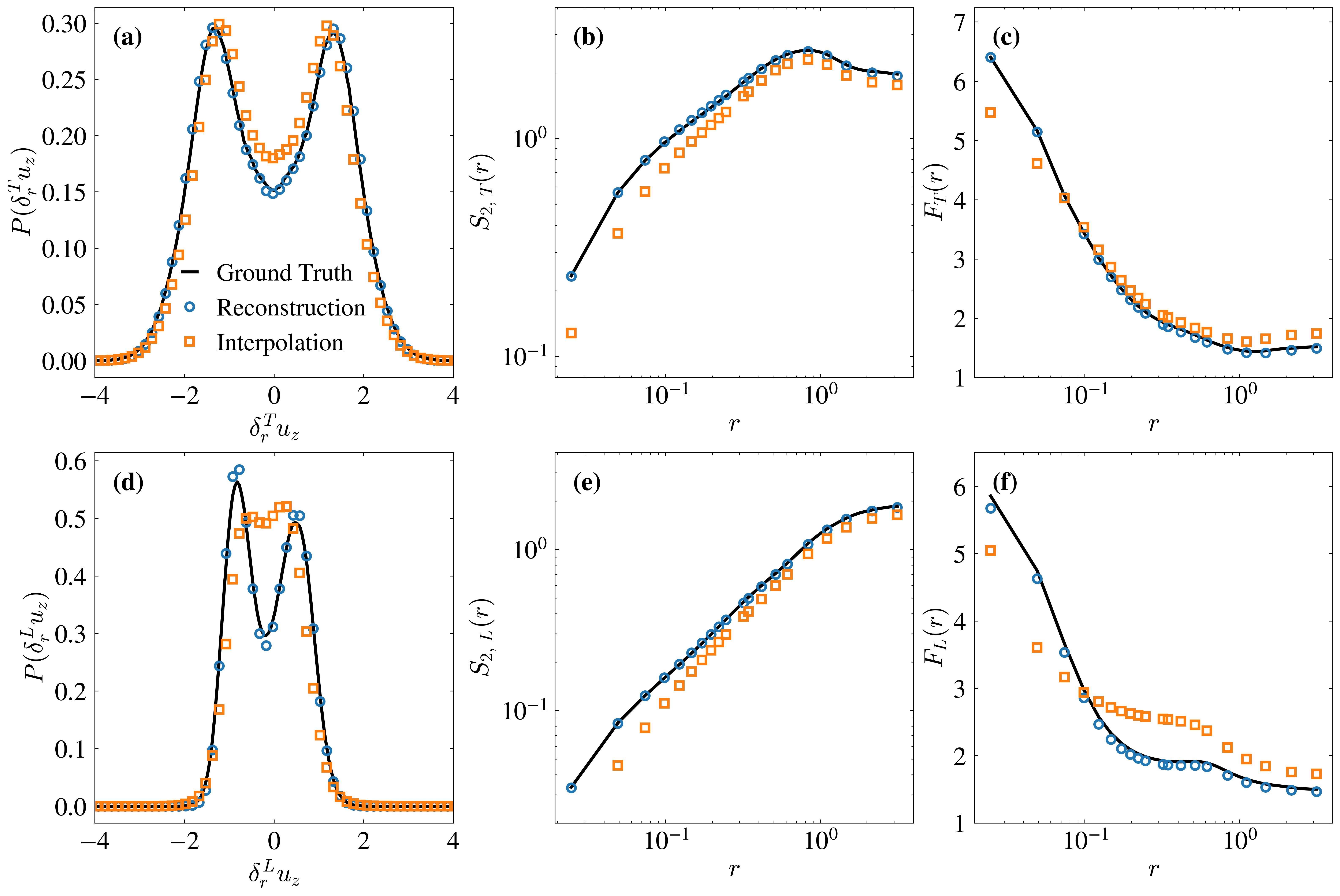}
\caption{\label{fig:reconstruction_statistics_across} PDFs of the transverse and longitudinal velocity increments $\delta_r^{(L,T)} u_z$ for $r \approx D/3$ (a,d), the corresponding transverse and longitudinal second-order structure functions (b,e), and the associated flatness (c,f). Only increments taken across the bubble–liquid interface are included.}
\end{figure*}
We now address the following experimentally relevant question: given knowledge of the bubble configuration and the velocity field in the liquid phase, can we reconstruct the flow inside the bubble?
To this end, we employ the training-free guidance method known as diffusion posterior sampling (DPS) \citep{Chung2024}, which was briefly discussed in section~\ref{subsec:tfg_dps}.
We remind the reader that DPS enables us to reuse the same diffusion model to generate samples conditioned on information available within a specified measurement region $\mathcal{M}$. 
The strength of the bias is governed by the guidance hyperparameter $\zeta$ (see Eq. \eqref{eq:backward_sample_conditional}).

In the following, the liquid phase ($\phi\leq 1/2$) is the measurement region $\mathcal{M}$.
The ground truth is the validation dataset {\tt R3a} for determining the optimal guidance strength and the testing dataset {\tt R3b} for subsequent comparisons.
 
To obtain the optimal guidance strength $\zeta^{\rm opt}$, we reconstruct the flow inside the bubbles for the $N_R=300$ slices in {\tt R3a}, using eight different values of $\zeta \in [32, 1300]$.
The normalized mean square difference between the ground truth (DNS flow fields) and the reconstructed fields in the bubble phase $M_i^b$ and the liquid phase $M_i^l$ is defined as: 
\begin{align}
    M_i^{b,l}(\zeta) = \frac{1}{N_R}\sum_{r=1}^{N_R}\frac{\int \dd^2\bx\,\phi_r^{b,l} (u_{r,i} - u_{r,i}^{\text{RE}}[\zeta])^2}{\int \dd^2\bx\, \phi_r^{b,l} u_{r,i}^2},
    \label{eq:nmse}
\end{align}    
where $\phi^b=\phi$,\, $\phi^l=1-\phi$, and $u_i^{\text{RE}}[\zeta]$ represents the reconstructed velocity field with guidance strength $\zeta$.  Fig.~\ref{fig:optimal_2d_guidance} displays the mean square error $M_i$ as a function of $\zeta$ for all three components of the velocity in both phases.
Although the minima occur at the same value of the guidance strength for the different velocity components within a phase, the vertical component exhibits a smaller mean square error.
This is likely due to the presence of coherent flow structures in the vertical velocity. In addition, since the guidance in Eq. \eqref{eq:backward_sample_conditional} is applied exclusively in the liquid phase, we have $M^l \ll M^b$.
Therefore, we choose the optimal guidance strength $\zeta^{\rm opt}=800$ as the value that minimizes the mean square error in the bubble phase.
We discuss the choice of the optimal guidance strength in section \ref{subsubsec:choose_opt_zeta}.

We now apply the diffusion model to reconstruct the flow inside the bubble phase, given the liquid-phase velocity field and the bubble positions for the dataset {\tt R3b} containing 3,200 $2d$ slices.
The plot in Fig.~\ref{fig:reconstruction_visual2d} presents a representative snapshot of the vertical velocity $u_z$ and its reconstruction using Eq. \eqref{eq:backward_sample_conditional}.
As a baseline, we additionally use the {\tt griddata} interpolation routine from the SciPy library \citep{2020SciPy-NMeth} to estimate the flow inside the bubbles (see the supplementary material~\citep{SM_self2026} for more details).
Compared with the naive interpolation method, the diffusion-model reconstruction more accurately captures the flow structures.

We stress that the reconstruction procedure is inherently statistical. Using $\zeta=\zeta^{\rm opt}$ together with the same liquid-phase velocity field and bubble arrangement, we obtain multiple realizations of the reconstructed vertical velocity.
In Fig.~\ref{fig:various_realizations_1dcut}, we show the corresponding one-dimensional cuts of the vertical velocity for the four bubbles displayed in Fig.~\ref{fig:reconstruction_visual2d}.
These plots clearly demonstrate that a typical reconstruction reproduces the velocity profile much more accurately than interpolation, which either overpredicts $u_z$ along the vertical cuts or introduces spurious high-frequency spatial oscillations in the horizontal direction.

We now investigate how accurately the diffusion model reproduces the statistical characteristics of velocity fluctuations in the bubble phase. In Fig.~\ref{fig:pdf_all_bubble_master}, we present the PDFs of the horizontal and vertical velocity components (and their derivatives) in the bubble phase for the test dataset (ground truth, {\tt R3b}), the reconstructed flow fields, and the interpolated flow fields.
The PDFs derived from the reconstructed flow are in close agreement with the corresponding ground-truth distributions. 
In contrast, in most cases the interpolation scheme consistently fails to represent both the core and the tails of the PDFs correctly.

To assess whether the reconstructed flow fields correctly reproduce the bubble–liquid correlations, we also evaluate the distribution of vertical velocity increments taken \textit{across} the interface. 
Specifically, we define horizontal (transverse) increments as $\delta^T_r u_z = u_z(h+r, z)-u_z(h, z)$ and vertical (longitudinal) increments as $\delta^L_r u_z = u_z(h, z+r)-u_z(h, z)$, where $h$ and $z$ denote the horizontal and vertical coordinates, respectively, and we only consider increments that traverse the bubble–liquid interface.

The panels in Fig.~\ref{fig:reconstruction_statistics_across} display velocity increment PDFs at $r\approx D/3$, the second-order structure functions $S_{2, (L,T)}=\langle (\delta_r^{(L,T)} u_z)^2 \rangle$, and the flatness $F=S_4/S_2^2$ for the reconstructed fields, which are indistinguishable from the corresponding statistics of the ground truth.
The interpolation scheme, however, not only fails to reproduce the core of the distribution, but also overestimates intermittency at intermediate scales while underestimating it at small scales.

\subsubsection{Choice of the optimal guidance strength $\zeta^{\rm opt}$}\label{subsubsec:choose_opt_zeta}
We now examine how sensitive $\zeta^{\rm opt}$ is to the sample size, as well as to our use of the mean square error (Eq.~\eqref{eq:nmse}) for determining it. 
We have verified that while minimizing the mean square error over $N_R=300$ slices leads to $\zeta^{\rm opt}=800$, reducing the sample size to $N_R=100$ yields the same optimal value.

We also remark that alternative metrics could be adopted to select $\zeta^{\rm opt}$, such as the Kullback–Leibler (KL) divergence between the distributions of $u_i$ in the ground-truth and reconstructed flow fields. 
The KL divergence is a measure of how much a probability distribution $P(X)$ differs from a model (approximating) distribution $Q(X)$, defined as follows~\cite{shlens_2014}:
\begin{equation}
    D_{KL}(P\|Q) = \sum_X P(X) \log \frac{P(X)}{Q(X)}.
\end{equation}
For the present case, we find that minimizing the KL divergence between $p^{\rm GT}(u_i)$ and $p^{\rm RE}(u_i)$ yields a slightly higher value of the optimal guidance strength, $\zeta\approx 1300$.
However, the resulting statistics of the reconstructed fields are similar in both the cases (not shown). 
Because the training objective minimizes the mean square error (see Eq. \eqref{eq:ddpm_loss}), we believe it is a natural criterion for identifying $\zeta^{\rm opt}$.
\begin{table*}       
  \centering 
  \begin{tabular}{cccccccc}
\textbf{Ga} &$T/\tau_\eta$ & $\delta t/\tau_\eta$ &\textbf{Number of Snapshots}& \textbf{Number of Samples}& $\eta/\Delta x$\\[3pt]\hline 
\tt $300$ & 67&1.49  &45&3,758& 2.76\\ 
\tt $1200$  &224&3.25 &69&5,956& 0.93\\ 
  \end{tabular}
  \caption{Details about the DNS dataset used for testing generalizability of the model trained at ${\rm Ga}=600$.
  The Bond number $\Bo\approx1.75$ and the number of bubbles $=12$ (volume fraction $\approx 3.2\%$) are the same as in runs {\tt R1-R3} in Table \ref{tab:dataset}.  
  Both simulations are with $512^3$ collocation points.
  For testing, we select 3,200 samples randomly from each of the datasets above.
  }
  \label{tab:dataset2}
\end{table*}

Finally, although we expect $\zeta^{\rm opt}$ to be fairly robust to moderate changes in flow parameters, substantially different flow regimes (e.g., Stokes flow) may lead to a different optimal guidance strength.

\subsection{Generalization for different $\Ga$}
\begin{figure*}
\centering
  \raisebox{6mm}{\includegraphics[width=0.27\linewidth]{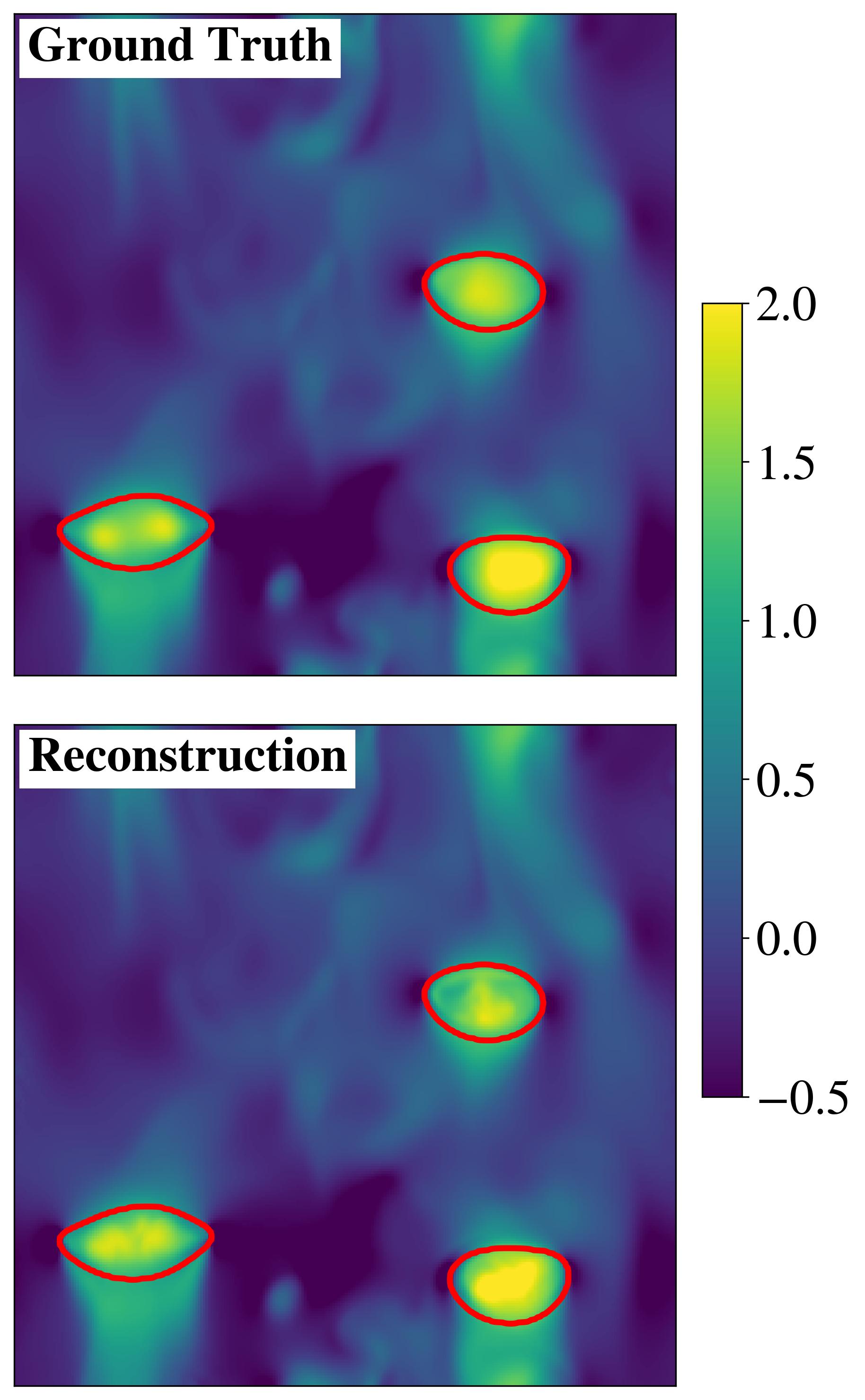}}
  \includegraphics[width=0.72\linewidth]{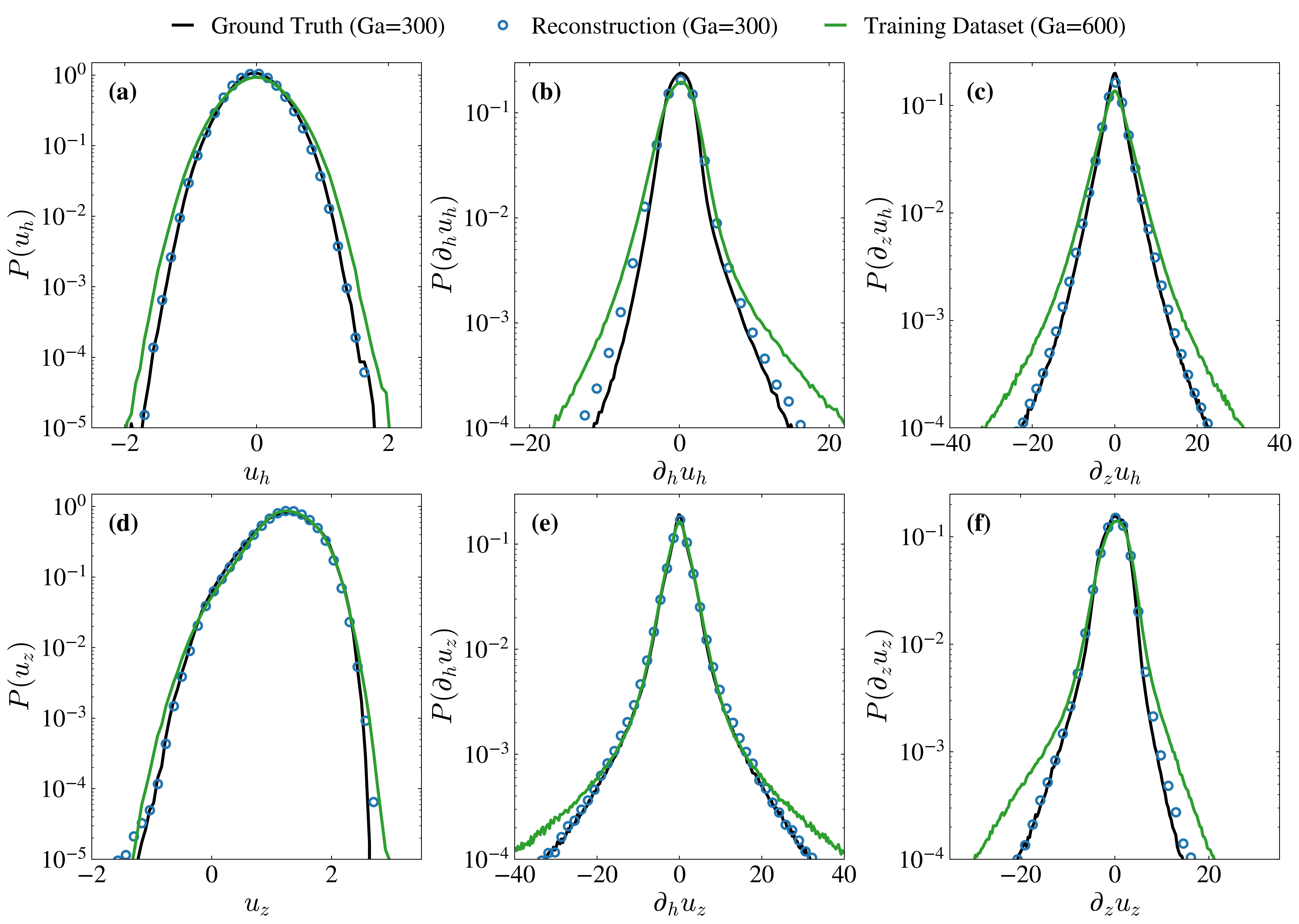}
\caption{Left: Pseudocolor plot of $u_z$ for the ground truth (top) and reconstruction (bottom) for ${\rm Ga}=300$.
Right: PDF of horizontal velocity (a) and its derivatives (b, c) and PDF of vertical velocity (d) and its derivatives (e, f) for the ground truth (${\rm Ga}=300$), reconstruction (${\rm Ga}=300$), and the training dataset (${\rm Ga}=600$) inside the bubble phase.
}\label{fig:generalization_ga300}
\end{figure*}
\begin{figure*}
\centering
  \raisebox{6mm}{\includegraphics[width=0.27\linewidth]{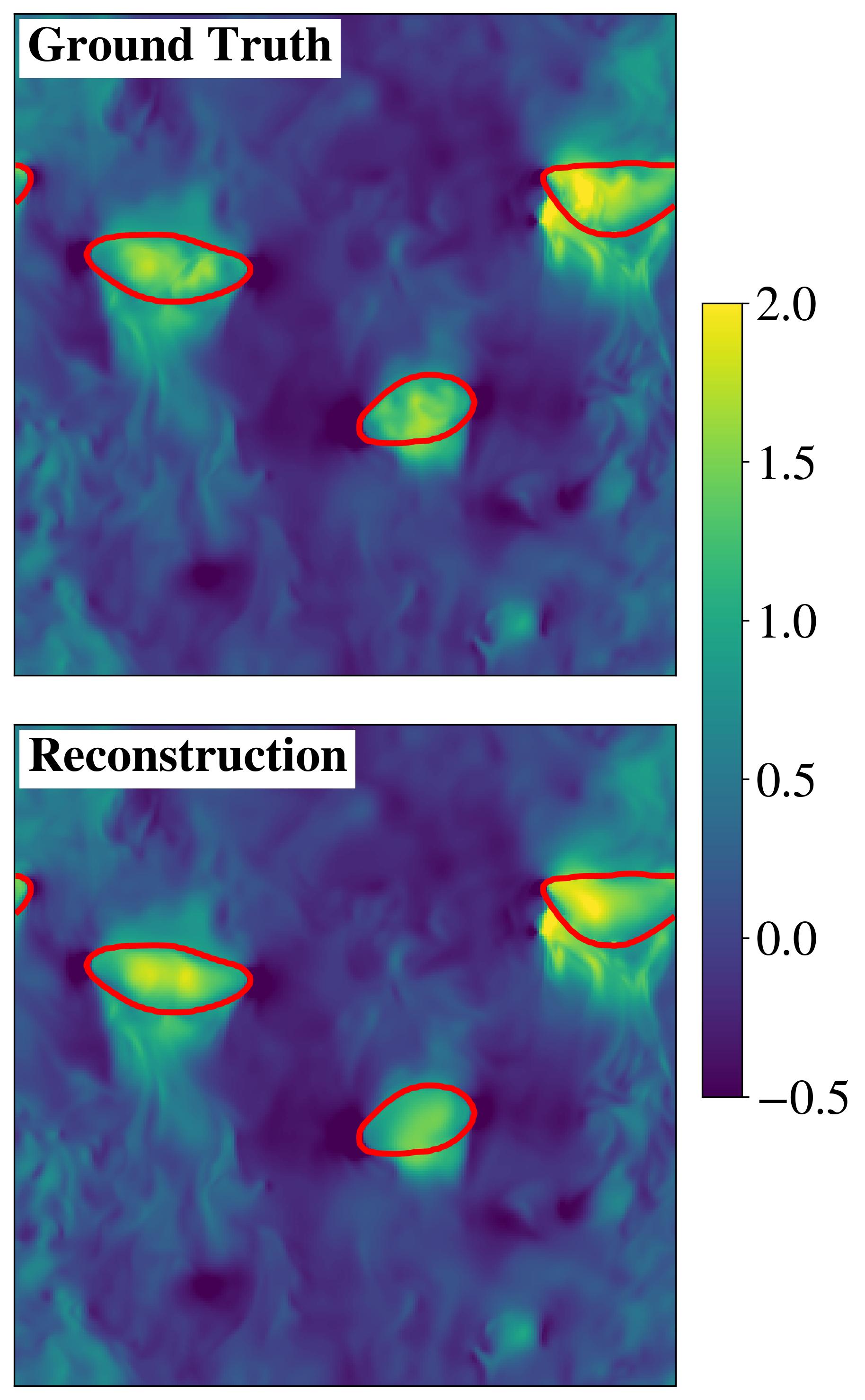}}
  \includegraphics[width=0.72\linewidth]{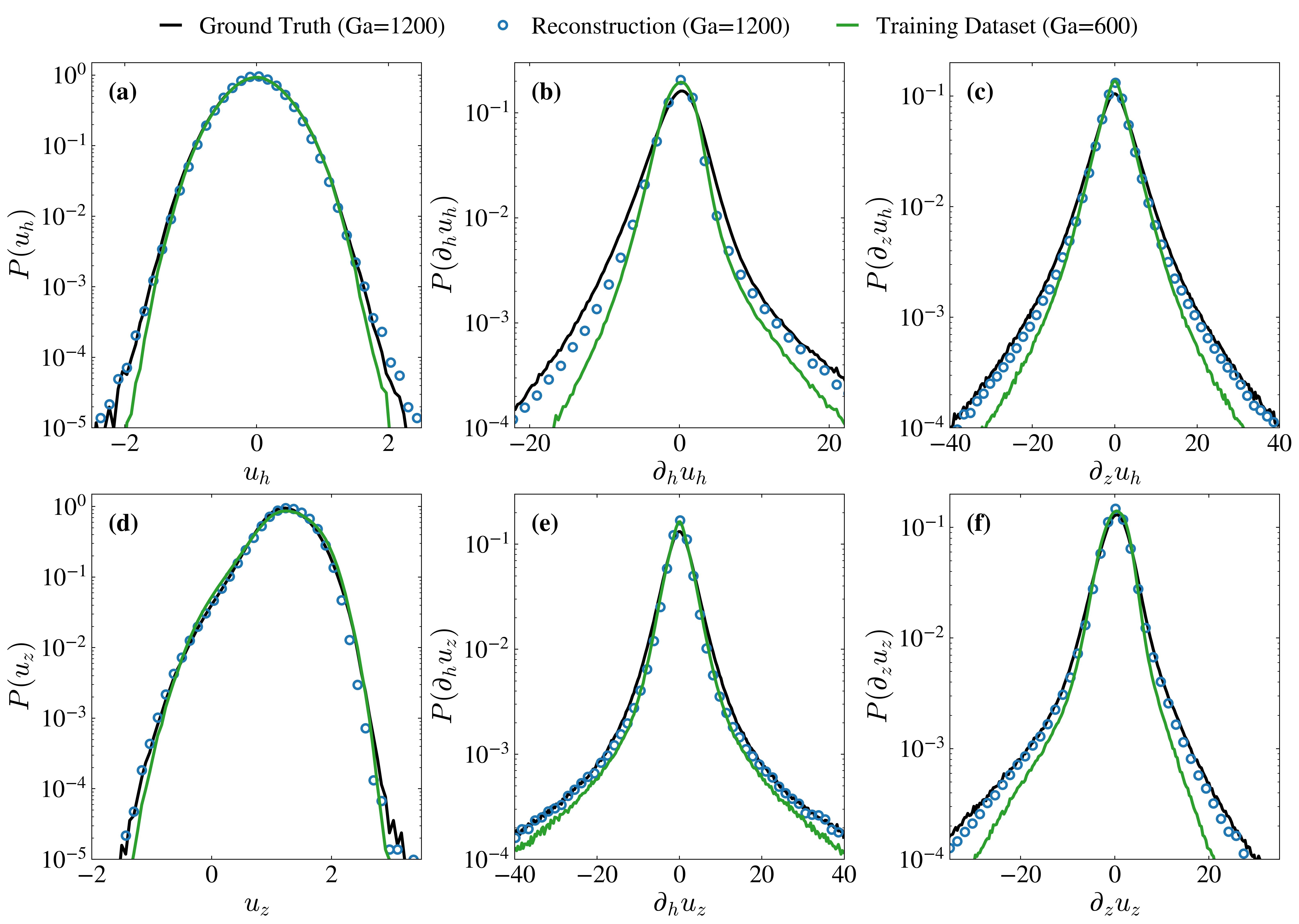}
\caption{Left: Pseudocolor plot of $u_z$ for the ground truth (top) and reconstruction (bottom) for ${\rm Ga}=1200$.
Right: PDF of horizontal velocity (a) and its derivatives (b, c) and PDF of vertical velocity (d) and its derivatives (e, f) for the ground truth (${\rm Ga}=1200$), reconstruction (${\rm Ga}=1200$), and the training dataset (${\rm Ga}=600$) inside the bubble phase.
}\label{fig:generalization_ga1200}
\end{figure*}
Thus far we have used a fixed Galilei number ${\rm Ga}\approx 600$ for both the model training and the flow reconstruction. We now extend our analysis to flow reconstruction within the bubble phase for datasets with smaller ${\rm Ga}\approx300$ and larger ${\rm Ga}\approx1200$  using the model trained at ${\rm Ga\approx 600}$.

The datasets for $\Ga\approx 300$ and $1200$ are prepared using the same numerical scheme and data acquisition procedure as discussed earlier in sections \ref{sec:dns_details} and \ref{subsec_data_acquire} respectively. The details about the datasets are provided in Table~\ref{tab:dataset2}.
To reconstruct the flow inside the bubble for different values of ${\rm Ga}$, we again set $\zeta^{\rm opt}=800$ and use the liquid-phase flow field ${\cal M}$ associated with the corresponding ${\rm Ga}$.

Representative visual comparisons between the ground-truth and reconstructed vertical velocity fields (Fig.~\ref{fig:generalization_ga300} and Fig.~\ref{fig:generalization_ga1200}) show that the reconstructed flows preserve characteristics of ${\rm Ga}=600$, the flow regime used for training. Specifically, the reconstruction appears slightly rougher than the ground truth at ${\rm Ga}=300$, and somewhat smoother than the ground truth at ${\rm Ga}=1200$.

To provide a more quantitative evaluation, in  Figs.~\ref{fig:generalization_ga300} and \ref{fig:generalization_ga1200} we examine the PDFs of the velocity fields and their derivatives within the bubble phase. As expected, the ground-truth PDFs at ${\rm Ga}=1200$ (${\rm Ga}=300$) show heavier (lighter) tails compared with those of the training dataset at ${\rm Ga}=600$. We then compare the PDFs computed from the reconstructed flow in the bubble phase against the ground truth. 
Notably, the reconstructed PDFs successfully reproduce the changes in the tail of the distributions. Nevertheless, for the in-plane derivatives of the horizontal velocity component, the tails are slightly overestimated (${\rm Ga=300}$, Fig.~\ref{fig:generalization_ga300}b) or underestimated (${\rm Ga}=1200$, Fig.~\ref{fig:generalization_ga1200}b). This aligns with the earlier observation that the reconstructed flow appears rougher or smoother depending on ${\rm Ga}$.

The statistical fidelity of the reconstructed flow fields to the ground truth (at least for the one-point statistics) is not entirely surprising since the reconstruction procedure is conditioned on the liquid flow information.
Even so, these results indicate that the reconstruction procedure is somewhat robust to changes in the Galilei number.
Likewise, we expect the model to perform reasonably well when other flow parameters are moderately varied.

With this, we complete our discussion of the two-dimensional reconstruction of the flow inside the bubble given the liquid velocity fields. In the next subsection, we show that the same model trained on the $2d$ dataset can also be used to reconstruct the $3d$ flow reasonably well.

\subsection{Three-dimensional ($3d$) flow reconstruction inside bubbles}\label{sec:3d_reconstruction}
\begin{figure*}[b]
\centering
  \includegraphics[width=0.45\linewidth]{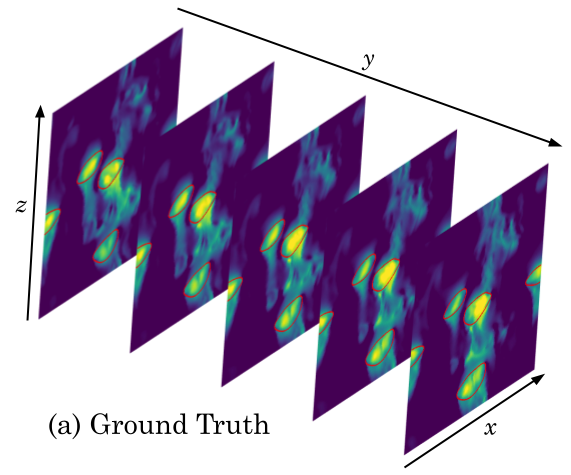}\,\,
  \includegraphics[width=0.45\linewidth]{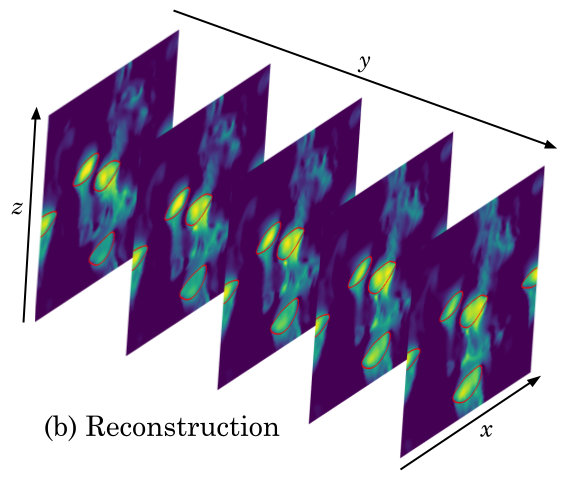}\,
\raisebox{4mm}{\includegraphics[width=0.06\linewidth]{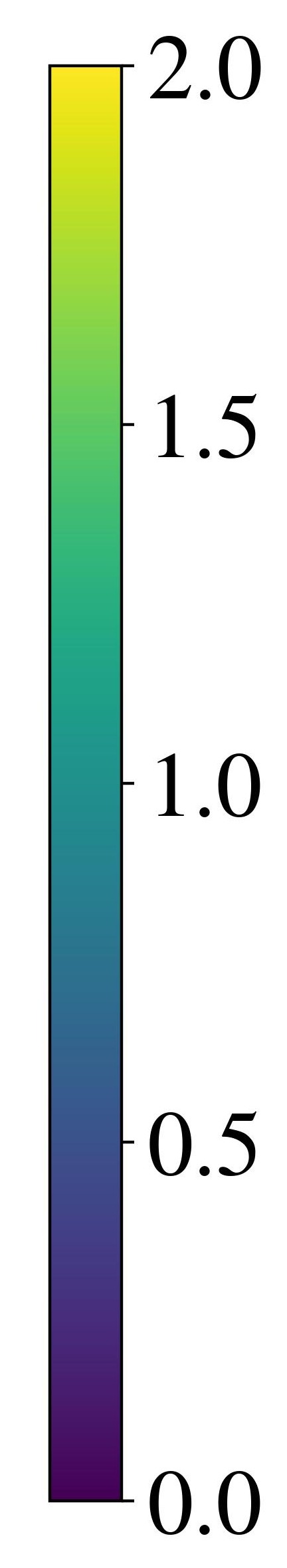}}
\caption{Visualization of a few representative slices of $u_z$ selected for the $3d$ reconstruction for the ground truth (a) and reconstruction (b). 
We have averaged the reconstructed flow fields over $60$ possible reconstructions for each slice, obtained by using different noise seeds.
Note that the horizontal direction is $x$, the vertical direction is $z$ and the out-of-plane direction is $y$.
}\label{fig:3dstacks_groundtruth_reconstruction}
\end{figure*}

Although the model is trained to reconstruct flow only in two-dimensional slices, we demonstrate below that stacking neighboring slices yields a plausible three-dimensional reconstruction of the flow within the bubble phase.
We select a single realization from run {\tt R3} and coarse-grain it onto a $256^3$ grid using a sharp spectral filter. For the test (ground-truth) data, we extract a sub-volume of size $256 \times 71 \times 256$ grid points, chosen such that at least one bubble lies fully within the domain.
Flow visualizations for a few representative slices in this sub-volume for the ground truth are shown in Fig.~\ref{fig:3dstacks_groundtruth_reconstruction}a.   
\begin{figure*}
\centering
  \includegraphics[width=0.45\linewidth]{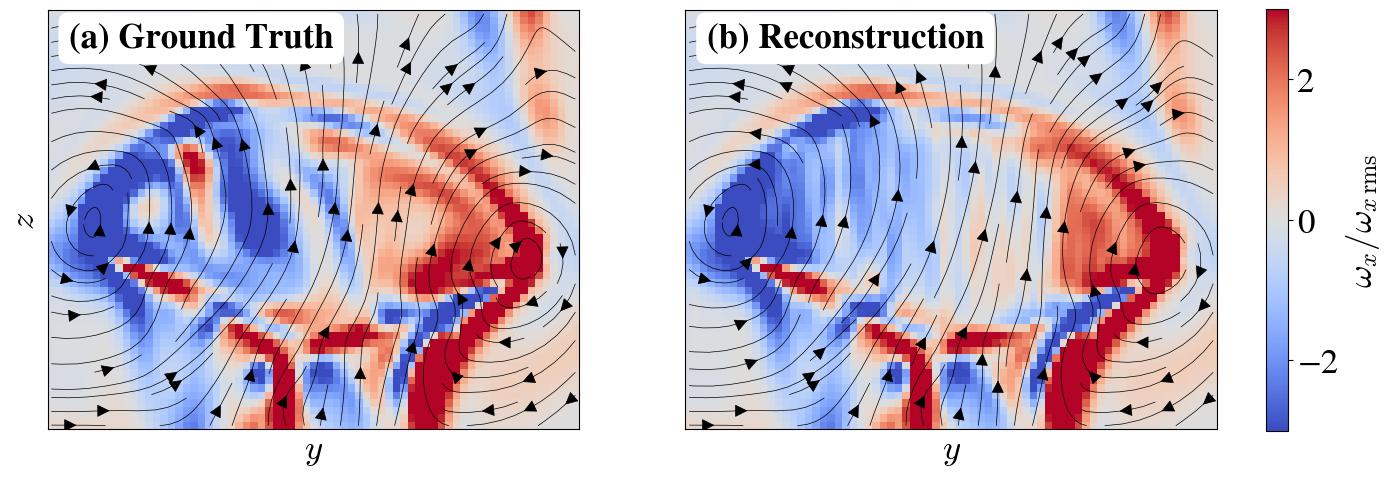}
    \includegraphics[width=0.25\linewidth]{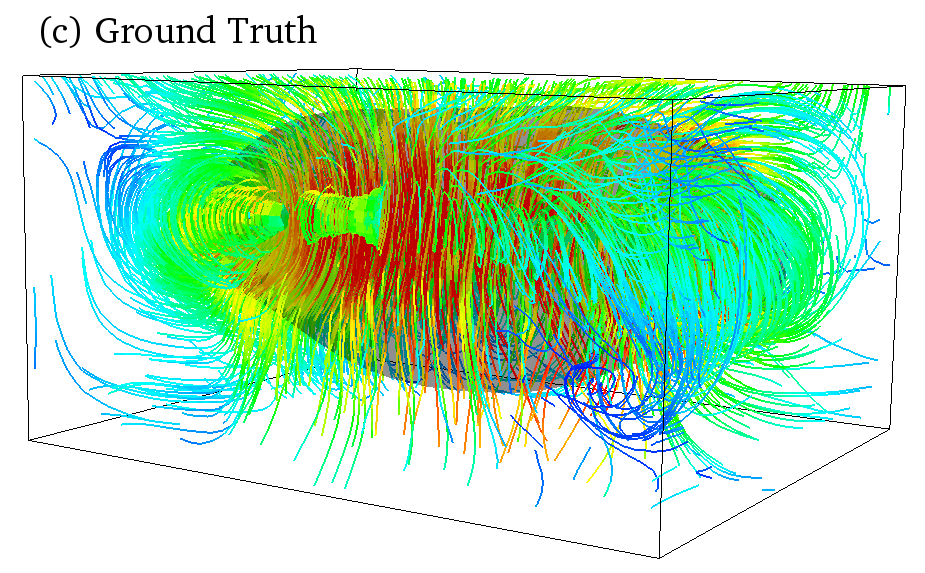}
  \includegraphics[width=0.25\linewidth]{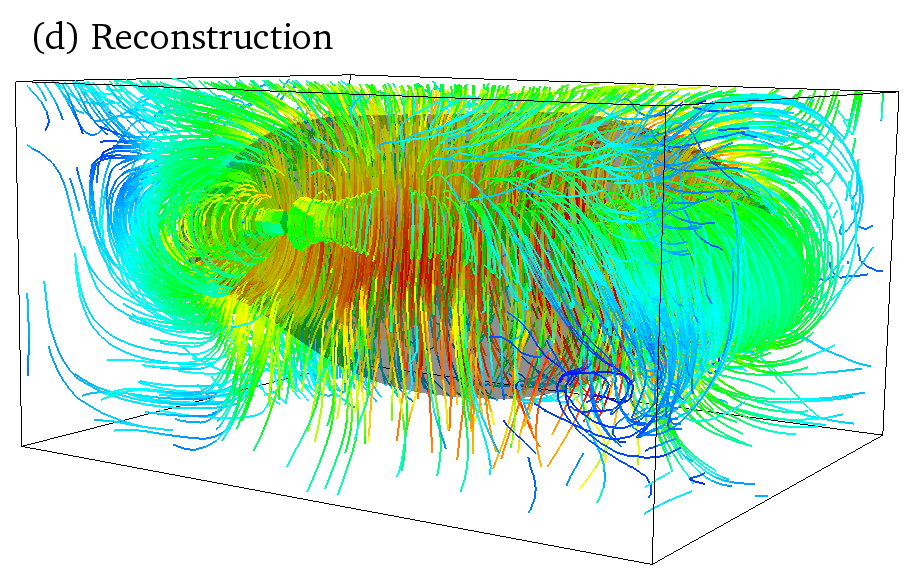}
\caption{\label{fig:3dreconstruction_xslice} (Left) Ground-truth (a) and reconstructed (b) vorticity fields $\omega_x$ in the $yz$ plane, shown together with the projected flow streamlines. (Right) Three-dimensional streamlines overlaid on the bubble surface for the ground truth (c) and the reconstructed flow (d).}
\end{figure*}
\begin{figure*}
\centering
  \includegraphics[width=0.99\linewidth]{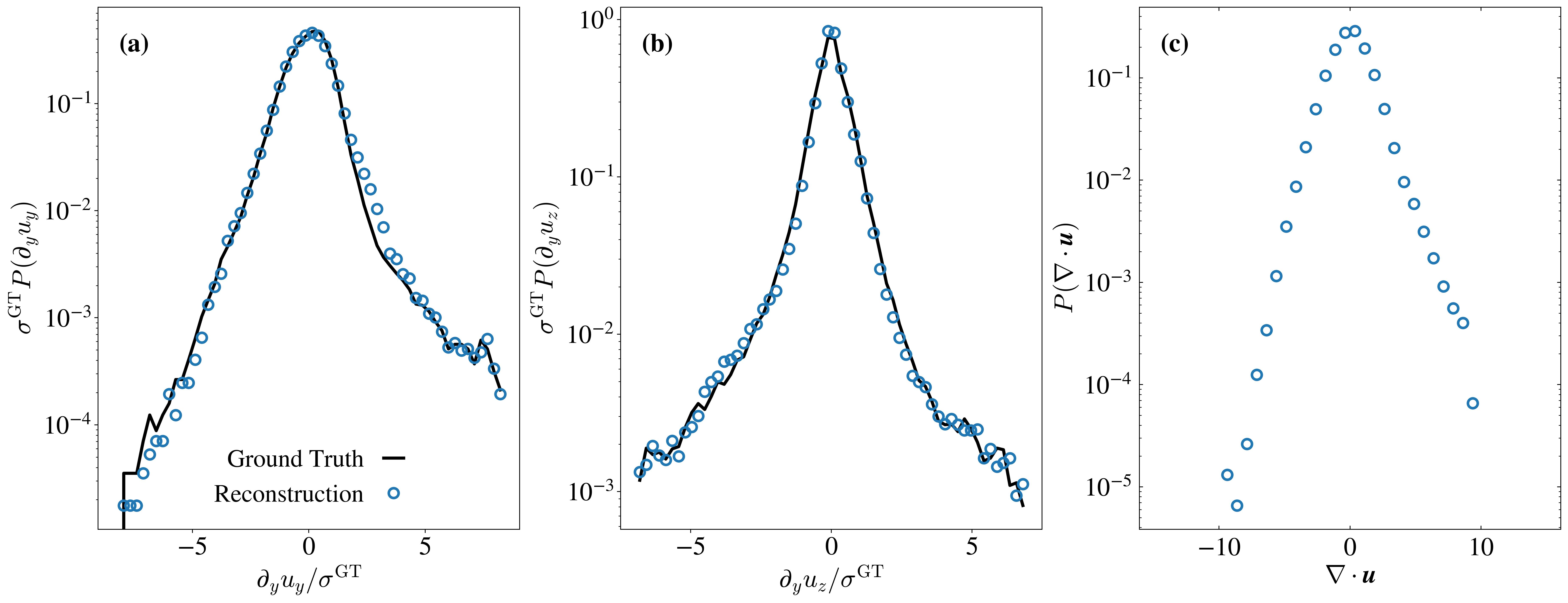}
\caption{\label{fig:3dreconstruction_result} PDF of (a) the out-of-plane derivative of $u_y$, (b) the out-of-plane derivative of $u_z$, and (c) the divergence of the velocity field. For (a) and (b) we have plotted the PDFs standardized with the ground-truth standard deviation.
}
\end{figure*}
Because the diffusion model is trained to reconstruct the flow within the bubble in the vertical ($xz$ and $yz$) planes, we reconstruct the flow separately for each of the $71$ $xz$ slices. 
However, this reconstruction procedure does not guarantee continuity of the velocity field in the $y$ (i.e., out-of-plane) direction. 
To mitigate this, we generate multiple ($60$) candidate flow reconstructions within the bubble phase for each $y$-slice and then take an ``average'' over these realizations, which smooths out strong small-scale fluctuations. The plot in Fig.~\ref{fig:3dreconstruction_xslice} shows the ground truth and the reconstructed $x$-component of the vorticity field $\omega_x=\partial_y u_z-\partial_z u_y$, and we observe that large-scale flow structures are well captured.

As a quantitative assessment of the $3d$ reconstruction, in Fig.~\ref{fig:3dreconstruction_result}a,b we present the PDFs of the derivatives $P(\partial_y u_y)$ and $P(\partial_y u_z)$ for the reconstructed flow in the bubble phase and observe that they closely match the ground truth.
Nevertheless, we emphasize that our naive $3d$ reconstruction does not  enforce pointwise incompressibility; the distribution $P(\nabla \cdot {\bf u}) \neq \delta(\bnabla\cdot\bu)$ (see Fig.~\ref{fig:3dreconstruction_result}c).

\section{Summary and Conclusion}
Our paper addresses the challenging problem of reconstructing the flow field within the bubble phase of a buoyancy-driven bubbly flow using diffusion models.
While AI/ML techniques have been previously used to reconstruct missing regions in homogeneous turbulence~\cite{SHU_2023,Li_atmos2024,Trepat_2026}, their application to multiphase flows has remained largely unexplored.
Such a setting is particularly challenging because, in bubbly flows, the statistical properties of the liquid and bubble phases are known to differ significantly \citep{Mercado_2010,ROGHAIR_2011,Pandey_Ramadugu_Perlekar_2020,PandeyMitraPerlekar2023,Innocenti_Jaccod_Popinet_Chibbaro_2021}.
Moreover, our study is also experimentally relevant, as direct flow measurements within the bubble phase are not feasible using optical diagnostics or direct in situ methods \citep{Prakash_2016,Sun_2017,RuiNi2024,ma2025}. 
Consequently, our proposed framework can be integrated with experiments to infer and analyze the flow dynamics inside bubbles.
While exact solutions are available for the flow field inside a bubble \citep{clift2005bubbles}, they apply only when the bubble remains almost spherical and the Galilei number is sufficiently low. In the present situation, however, bubble deformations are significant and the Galilei number is moderately large, so these analytical solutions cannot be applied directly.

The training and test datasets are obtained from fully resolved three-dimensional ($3d$) direct numerical simulations (DNS) of buoyancy-driven bubbly flows in the pseudo-turbulent regime ($\Ga \approx 600$) \cite{Pandey_Ramadugu_Perlekar_2020,Pandey_Mitra_Perlekar_2022}.
Since training the model directly on the complete $3d$ fields is computationally prohibitive, we instead train the diffusion model on two-dimensional vertical slices, each selected so that it is intersected by at least one bubble. 
We then use the diffusion posterior sampling (DPS) framework to infer the flow inside the bubbles, conditioned on the known liquid-phase velocity field.
We show that the model not only reproduces the large-scale flow patterns but also captures the statistics of the velocity field, its gradients, and two-point correlations with high accuracy. Furthermore, diffusion-based reconstruction substantially outperforms a smooth spline interpolation approach \citep{scipy_cloughtocher2dinterpolator}. We further demonstrate that the model trained at ${\rm Ga}=600$ remains robust to changes in the Galilei number and can reproduce the modifications of the flow within the bubble even for the two markedly different cases ${\rm Ga}=300$ and ${\rm Ga}=1200$.

Finally, we consider reconstruction of the three-dimensional flow within a bubble by stacking adjacent two-dimensional reconstructed slices.
Because the reconstruction is statistical in nature, continuity of the velocity field across slices is not inherently guaranteed.
To alleviate this, we generate multiple reconstructions ($60$) of the bubble-phase flow for each slice and then average them to obtain a smoother, more coherent flow field.
The resulting $3d$ bubble-phase flow closely reproduces the ground truth.
Furthermore, the statistics of the out-of-plane derivatives show good agreement with those obtained from the ground-truth data.
In conclusion, we demonstrate for the first time the feasibility of reconstructing the flow within the bubble phase using a guided diffusion model.
As outlined previously, the methodology and findings are of direct experimental relevance.
A natural extension of the present work would involve explicitly incorporating the bubble–liquid interfacial boundary conditions and the incompressibility constraint \cite{genuist_2026} into the diffusion model, with the aim of further enhancing the fidelity and accuracy of $3d$ flow reconstruction.
\begin{acknowledgments}
H.N. thanks Vikash Pandey for many helpful discussions on the numerical methods. H.N. and P.P. acknowledge various discussions with Vikash Pandey and Dhrubaditya Mitra on bubbly flows.\\
\textit{Funding:} This work was supported by the European Research Council (ERC) under the European Union's Horizon 2020 research and innovation program Smart-TURB (Grant Agreement No. 882340) and by the Italian Ministry of University and Research (MUR) - Fondo Italiano  per la Scienza (FIS2) - 2023 Call, Project DeepFL CUP : E53C24003760001, and  FARE programme (No. R2045J8XAW).
H.N. acknowledges support from the Smart-TURB (ERC) grant for his visit to the University of Rome, Tor Vergata, for this work.
H.N. and P.P. acknowledge support from the Department of Atomic Energy (DAE), India under Project Identification No. RTI 4007, and DST (India) Project No. MTR/2022/000867.
Simulations for obtaining the DNS datasets were primarily performed on the Kohinoor 5 HPC cluster (TIFR-H).
The training for the diffusion model was performed on Hydrosoft (University of Rome, Tor Vergata), with some additional testing on the Turing workstation (TIFR-H).
\end{acknowledgments}

\end{document}



\title{Supplementary Material for \\``Gappy Reconstruction of Bubbly Flows by Guided Diffusion Models''}
\author{Hridey Narula}
\email{hrideynarula@tifrh.res.in}
\affiliation{Tata Institute of Fundamental Research, Gopanpally, Hyderabad 500046, India}
\author{Tianyi Li}
\author{Michele Buzzicotti}
\author{Luca Biferale}
\affiliation{Department of Physics and INFN, University of Rome “Tor Vergata”, Via della Ricerca Scientifica 1, Rome, 00133, Italy}

\author{Prasad Perlekar}
\email{perlekar@tifrh.res.in}
\affiliation{Tata Institute of Fundamental Research, Gopanpally, Hyderabad 500046, India}
\date{\today}
\maketitle
\section{Diffusion Model Details\label{app:diffusion_models}}
\subsection{Variance Schedule}
The variance schedule $\beta_t$ is the strength of the noise at step $t$ and thus controls the amount of information lost at each step.
The Langevin update from the noisy image at step $t-1$ to $t$ is:
\begin{equation}\label{eq:langevin_noise}
    \mathbf{X}_t = \sqrt{1-\beta_t}\mathbf{X}_{t-1} + \sqrt{\beta_t}\boldsymbol{\epsilon},
\end{equation}
where the noise $\boldsymbol{\epsilon}$ is drawn from $\mathcal{N}(\mathbf{0}, \mathbf{I})$. 
The variance schedule $\beta_t$ is taken to be a simple function of step $t$, two common choices are linear and tan-hyperbolic functions \citep{Ho_2020,Li_nmi2024,Li_atmos2024,Li_ijmpf2024,Li_comm2025}.

\begin{figure}
\centering
  \includegraphics[width=0.4\linewidth]{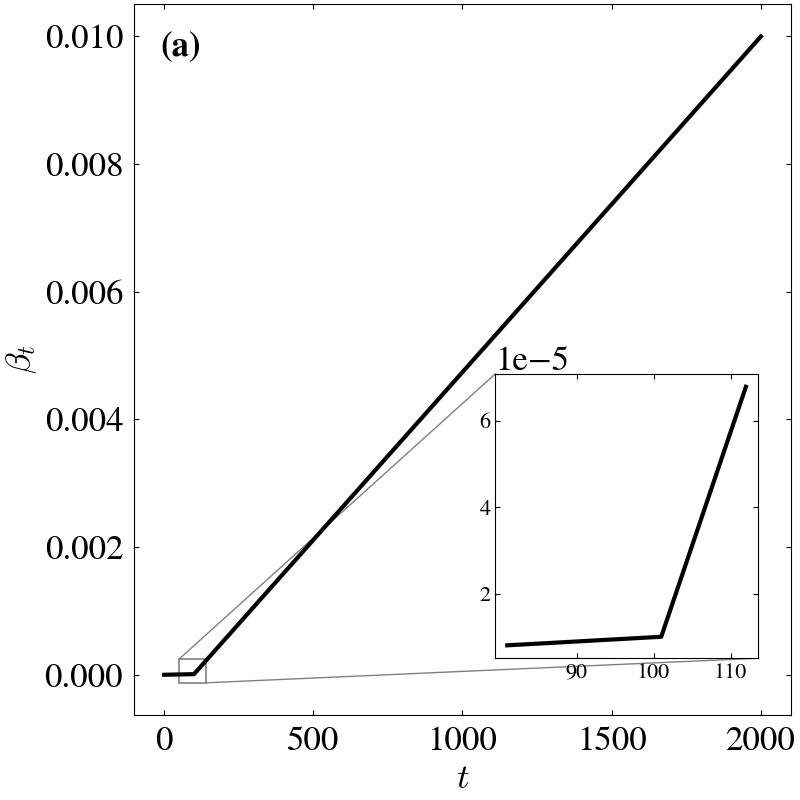}
  \includegraphics[width=0.4\linewidth]{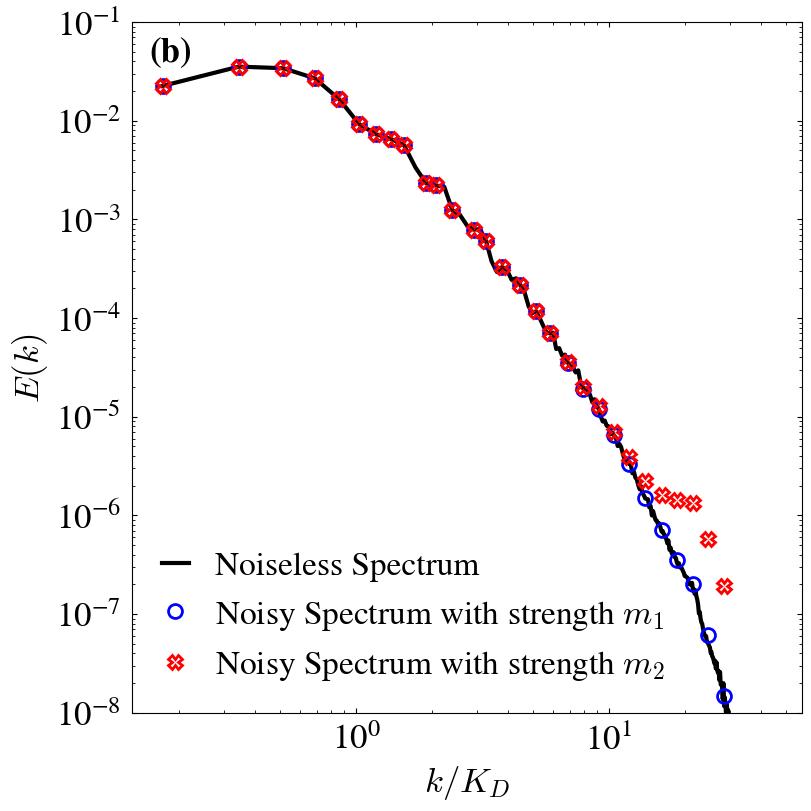}
\caption{(a): Variance schedule $\beta_t$ as a function of the noising step $t$.
The schedule is piecewise linear with slopes $m_1$ ($t<100$) and $m_2$ ($t\geq100$).
The inset is a zoom in near $t=100$ to show the change in slope.
(b): Energy spectrum of (i) the noiseless (clean) image (black solid line), and the noisy velocity field after a single update eq.~\eqref{eq:langevin_noise} with noise strength (ii) $m_1$ (blue circles) and (iii) $m_2$ (red crosses). The spectrum is for a single randomly selected $2d$ velocity slice from the training dataset.
\label{fig:noise_schedule}}
\end{figure}

For the present work, we use a ``bilinear'' (piecewise linear) schedule defined as follows (with the step $t\in[1, T]$):
\begin{equation}
    \beta_t =     \begin{cases}
        \beta_{\tau}+(t-\tau)m_2 &t\geq\tau\\
        \beta_1+(t-1)m_1 & t< \tau
    \end{cases}
\end{equation}
where $m_1$ and $m_2$ are the slopes for $t<\tau$ and $t\geq\tau$ respectively.
We choose $\beta_1=10^{-7}$, $\beta_\tau=10^{-5}$ and $\beta_T=10^{-2}$ with $T=2000$ and $\tau=100$. These parameters fix $m_1\approx10^{-7}$ and $m_2\approx5.3\times10^{-6}$.
The variance schedule is plotted in Fig.~\ref{fig:noise_schedule}a.

We have used this bilinear schedule because a linear schedule with slope $m_2$ destroys the small-scale fluctuations too quickly.
This is apparent in Fig.~\ref{fig:noise_schedule}b where we have plotted the energy spectrum of the noiseless velocity field along with the energy spectrum of the noisy velocity field after a single step of Eq. \eqref{eq:langevin_noise} with noise strength $m_1$ and $m_2$.

While in principle we can use a linear schedule with the smaller slope $m_1$, this would require many more noising diffusion steps, making the training slower.
Hence, we use a bilinear schedule with a smaller initial slope $m_1$ and a larger slope $m_2$ after the first $100$ steps.

\subsection{Model Architecture}
\begin{figure*}
\centering
  \includegraphics[width=0.99\linewidth]{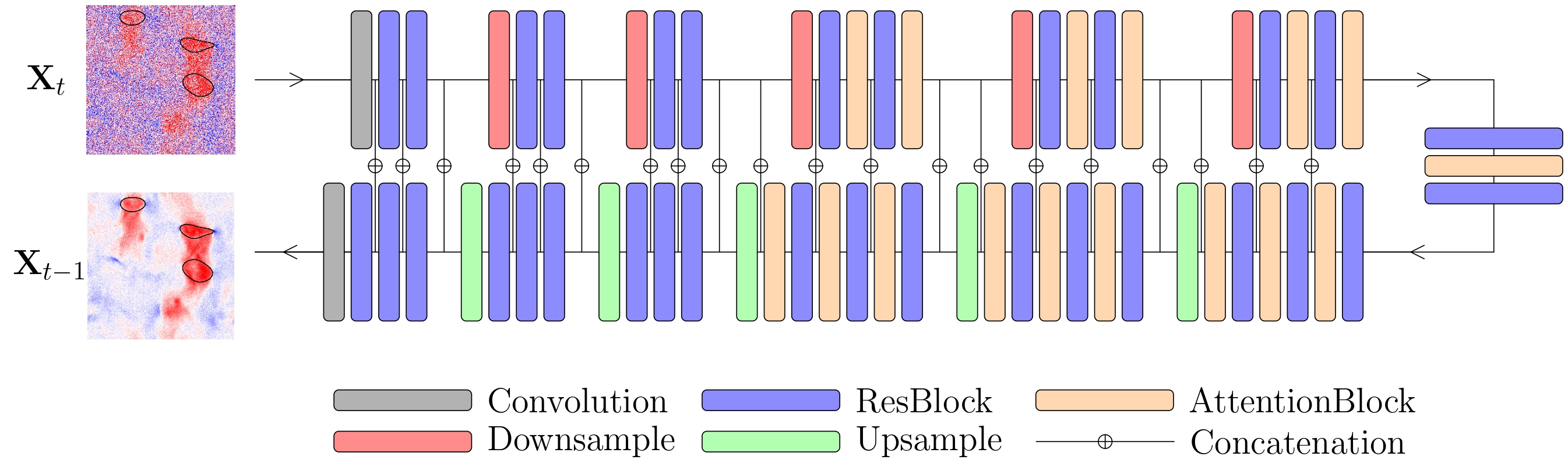}
\caption{Schematic for the U-Net architecture used in the present work.
The input is the noisy image $\mathbf{X}_t$ at step $t$ which passes first through the down-sampling (encoder) stack and then the up-sampling (decoding) stack, with the output being the slightly less noisy image $\mathbf{X}_{t-1}$.
Each down-sampling stage decreases the resolution by a factor of $2$.
The starting resolution is $256$; attention is applied at the last three stages with the resolution $32$, $16$ and $8$ respectively.
The down-sampling and up-sampling stacks are linked via skip connections (concatenation).
\label{fig:unet_architecture}}
\end{figure*}
We utilize a U-net architecture \citep{Ronneberger_2015} for the diffusion model.
The model architecture is schematically shown in Fig.~\ref{fig:unet_architecture}.
There are six stages in both the down-sampling (encoding) and up-sampling (decoding) stack.
Both the stacks are connected by skip connections.
The residual blocks are configured with channels $[C,C,2C,2C,4C,4C]$ where $C=256$.
Attention blocks \citep{Vaswani_2023} are applied at the last three stages, that is, at the resolutions of $32,\, 16,\,\text{and}\, 8$.
The number of head channels used is $64$.
Further, the AdamW optimizer \citep{Loshchilov_2019} is used in training. We fix the learning rate to $5\times10^{-5}$.
Total iterations are $4.9\times10^5$ with a batch size of $4$ on eight NVIDIA A100 GPUs, which took approximately $96$ hours.
The total number of training dataset samples was $16944$ so that each epoch corresponds to $16944/(4\times8)\approx529$ iterations.
Finally, to generate new samples we use an exponential moving average (EMA) method with a decay rate of $0.9999$.
More details about the model architecture can be found in the previous works \citep{Li_nmi2024,Li_atmos2024,Li_ijmpf2024,Li_comm2025}.

\section{Interpolation Details\label{app:interpolation_details}}
For $2d$ interpolation we have used the scipy function {\tt griddata} \citep{2020SciPy-NMeth,scipy_cloughtocher2dinterpolator}.
{\tt Griddata} is a general function to interpolate unstructured grid data.
The function takes as input the list of coordinates and the known function values at those points (for us, the liquid region) and can then interpolate the data on arbitrary points (for us, the bubble region).
We use the `cubic' mode which internally calls the {\tt CloughTocher2DInterpolator} function from the scipy library.
The resulting interpolant is a piecewise cubic and continuously differentiable function.
For more details, we refer the reader to the webpage~\citet{scipy_cloughtocher2dinterpolator} and references therein.
Finally, we note that to explicitly take care of masked points near the boundary, we use periodic boundary conditions to tile the data before supplying it to the griddata function.
For the one-dimensional cuts of two-dimensional slice, we also did a cubic spline interpolation, however, it was observed to be somewhat worse than the full two-dimensional interpolation and hence not shown.

\section{Reconstruction of Eulerian Time Series\label{app:1d_time}}

Finally, we also show that the diffusion models can also be used for temporal reconstruction of velocity signals.
As mentioned in the introduction, experimental techniques based on the material properties of fluids cannot obtain the flow velocity in both phases accurately.
Consider the Eulerian measurement of the velocity signal $\bu(t)$ at some fixed location $\bx_0$. 
When a bubble passes this point, the measurement records a spurious signal owing to the difference in the material properties (heat capacity, refractive index etc.) of the bubble and the liquid phases.
In such cases, careful signal processing is required to remove/smooth the corrupt part of the signal.
This can be done, for example, by application of a smoothing operator \citep{LanceBataille1991}, thresholding either the amplitude or the derivative of the amplitude of the signal \citep{ZENIT_KOCH_SANGANI_2001,Mercado_2007,Sun_2017}, pattern recognition \citep{RENSEN_LUTHER_LOHSE_2005}, using phase-sensitive temperature anemometry \citep{Mercado_2010,ZENIT_2013,Prakash_2016} or by restricting measurements to the wake behind the bubble swarm only \citep{RIBOUX_2010,Ravisankar_Zenit_2024}.
Naturally, while such methods can accurately identify the part of the signal where bubbles interact with the probe and thus remove the `corrupt signal', they miss out on the `true' velocity signal $\bu(t)$.

\begin{figure}[b]
	\centering
	\includegraphics[width=0.35\linewidth]{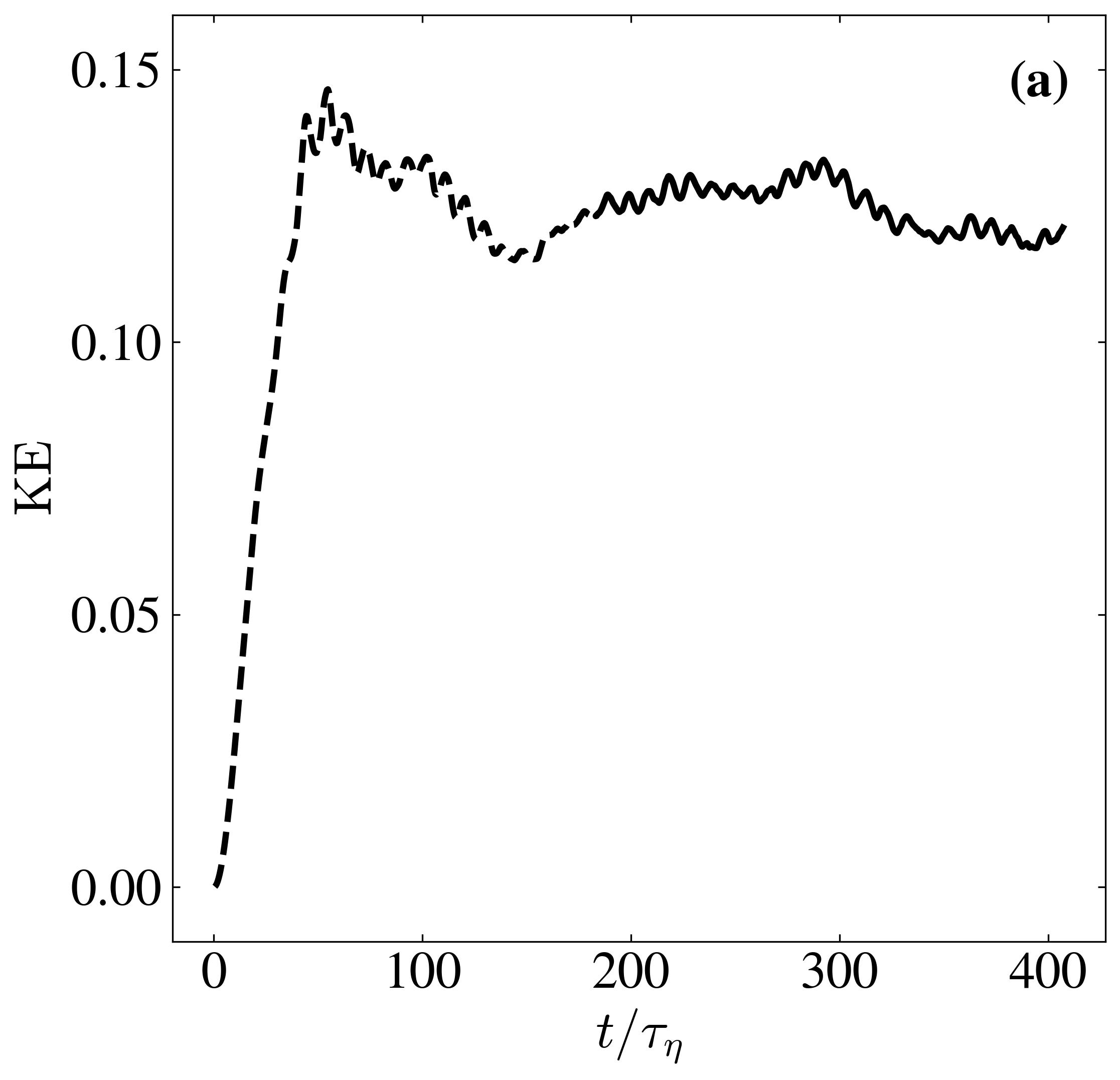}
	\raisebox{.4cm}{%
		\includegraphics[width=0.35\linewidth]{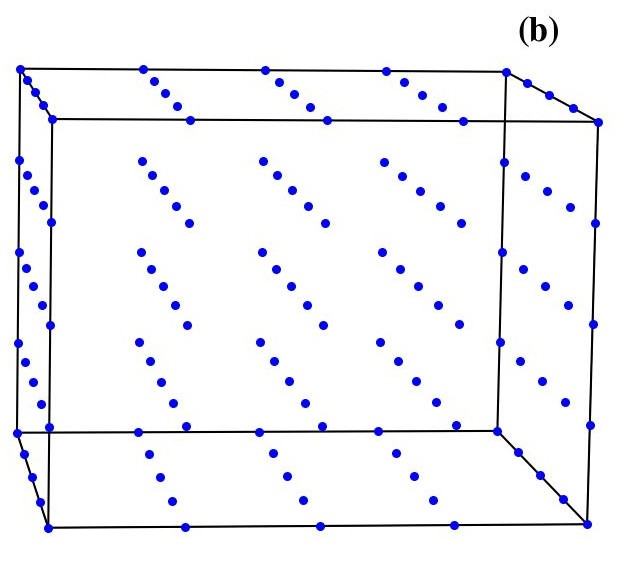}
	}
	\caption{(a): Kinetic energy time series of the DNS run used to obtain the Eulerian velocity time signals. Solid line indicates the region from which data is extracted.
		(b): An indicative selection of the Eulerian points where the velocity time signals are recorded.
		\label{fig:setup_1d}}
\end{figure}

We now show that diffusion models can be used to successfully reconstruct the Eulerian time signals $\bu(\bx_0, t)$, during the window where the bubble is present at the location $\bx_0$, provided the signal is known at all other times.
We use a fourth independent run with the same parameters ($\Ga=600$, $\At=0.04$, $\Bo=1.75$) as for runs {\tt R1-R3} but with $256^3$ collocation points.
The kinetic energy time series is shown in Fig.~\ref{fig:setup_1d}a.
We use this to obtain $70^3$ Eulerian time signals which are split into a $90:10$ training/testing dataset.
The probes are uniformly distributed in the box; a subset of the same is shown in Fig.~\ref{fig:setup_1d}b.
The signal length in the statistically steady state is about $186\tau_\eta$ with successive values separated by approximately $\tau_\eta/11$ (where $\tau_\eta=\sqrt{\nu/\varepsilon}$ is the Kolmogorov time scale) so that the time series has 2048 points.

\begin{figure}
	\centering
	\includegraphics[width=0.4\linewidth]{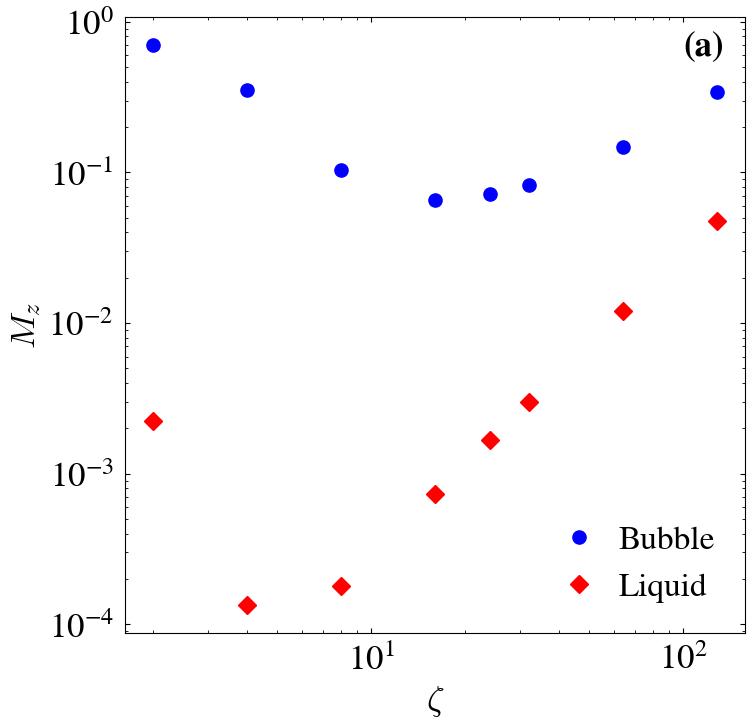}
	\includegraphics[width=0.4\linewidth]{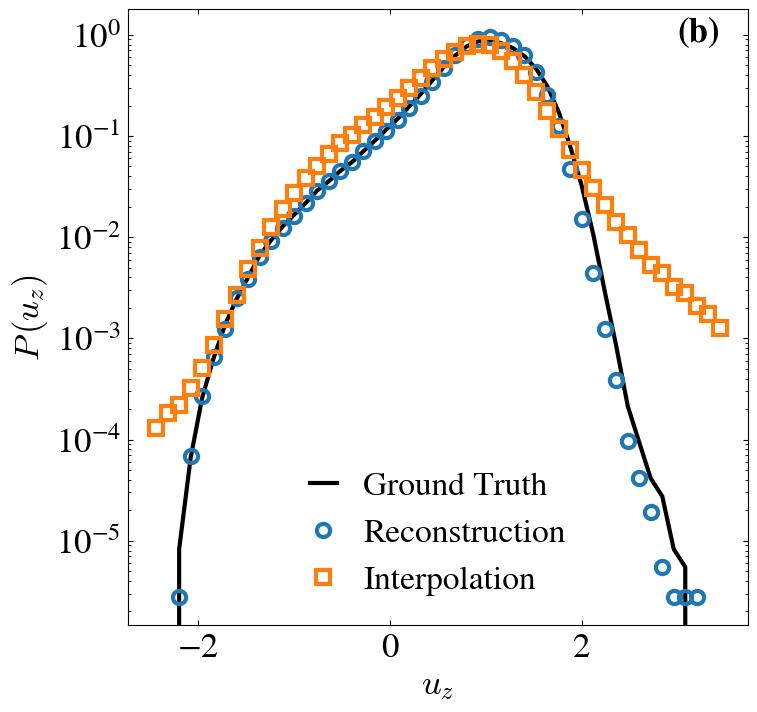}
	\caption{(a): Normalized mean square error $M_z$ for the vertical velocity $u_z$ in the bubble and liquid phases as a function of the guidance strength $\zeta$.
		(b): Normalized pdf of $u_z$ in the bubble phase for the ground truth, reconstruction and interpolation.
		Both are for 1$d$ temporal training.\label{fig:results_1d}}
\end{figure}

\begin{figure}
	\centering
	\includegraphics[width=0.4\linewidth]{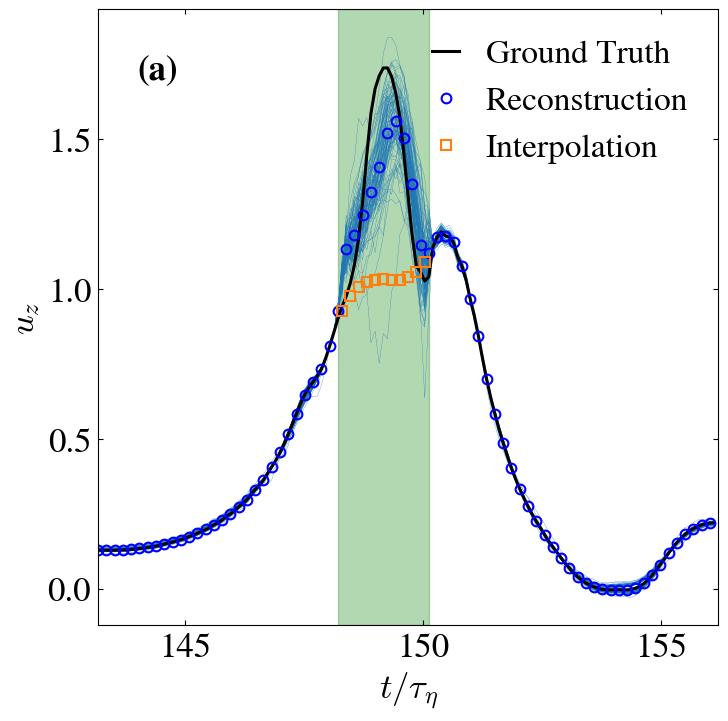}
	\includegraphics[width=0.4\linewidth]{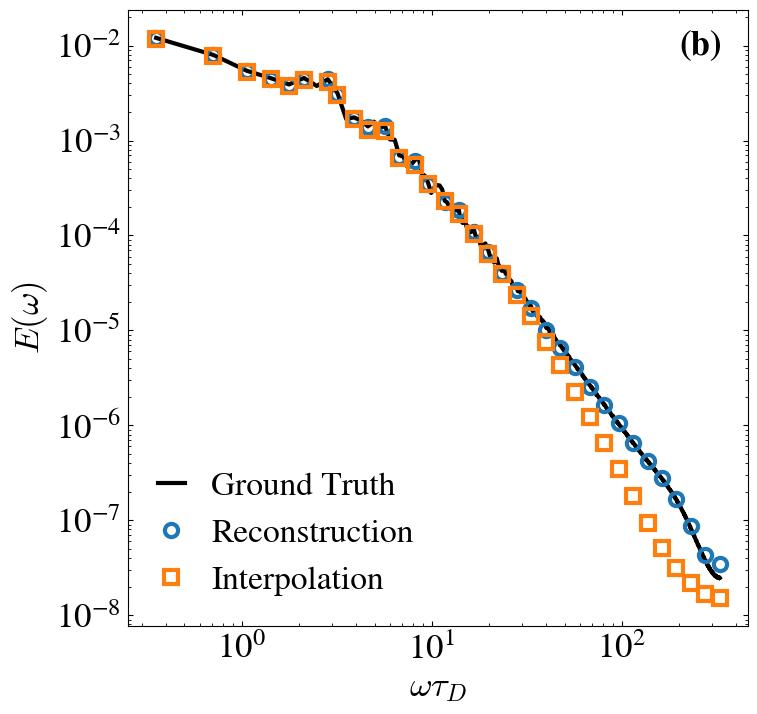}
	\caption{
		(a): A zoom in of the $u_z(t)$ signal showing various possible reconstructions as well as filling by cubic interpolation. The shaded regions indicate the window where a bubble crosses the Eulerian point.
		(b): The energy spectrum in the frequency space $E(\omega)$ with $\tau_D=D/u_{\rm rms}$ being the bubble time-scale. \label{fig:results_1d_2}}
\end{figure}

\begin{figure*}
	\centering
	\includegraphics[width=0.9\linewidth]{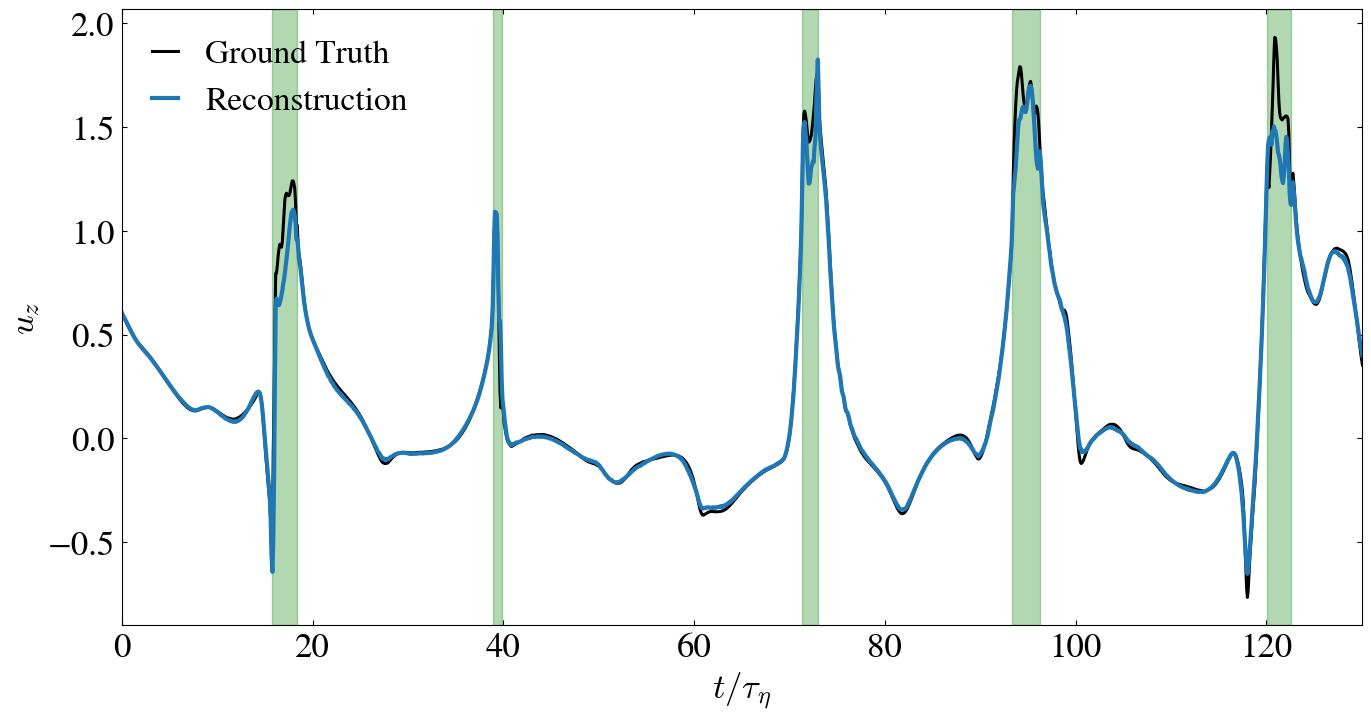}
	\caption{A longer time series for the ground truth as well as one possible reconstruction of the velocity signal $u_z$. The shaded regions mark the windows where a bubble crosses the Eulerian point.\label{fig:results_1d_full}}
\end{figure*}

We train the diffusion model on two channels: the indicator function $\phi$ and the vertical velocity $u_z$.
The model parameters are the same as those used in the earlier study on gappy reconstruction of Lagrangian turbulent signals \cite{Li_comm2025}.
We train the model for $250,000$ iterations using $8$ GPUs with a per GPU batch size of 128, taking about $40$ hours.
We again first find the optimal guidance strength as shown in Fig.~\ref{fig:results_1d}a.
Since the normalized mean square error in the velocity component $u_z$ is larger in the bubble phase, we minimize it by choosing the optimal guidance strength to be 16.
The plot of the pdf of the velocity signal in the bubble phase is shown in Fig.~\ref{fig:results_1d}b, where we also plot the result obtained using a simple cubic spline interpolation from the scipy library \citep{2020SciPy-NMeth}.
Since the interpolation scheme cannot reliably work for the edge points, for calculations of all the statistics we drop $100$ points from each end of the time-series.
In Fig.~\ref{fig:results_1d_2}a, we show a zoom-in of the time series of $u_z$ along with the reconstruction and cubic interpolation. Further, the temporal spectrum of the reconstructed signal as shown in Fig.~\ref{fig:results_1d_2}b agrees quite well with the ground truth spectrum.
Finally, we also show a longer time series of $u_z$ for both the ground truth and the reconstruction in Fig.~\ref{fig:results_1d_full}.

We thus conclude that diffusion models can also be used for reconstruction of gappy time signals, even when the statistics of the masked region (bubble) are different from the unmasked region (liquid).

